\documentclass[gmd, manuscript]{copernicus}

\usepackage{subcaption}

\begin{document}

\nolinenumbers

\title{ML-IAM v1.0: Emulating Integrated Assessment Models With Machine Learning}


\Author[1]{Yen}{Shin}
\Author[1]{Changyoon}{Lee}
\Author[2]{Eunsu}{Kim}
\Author[1]{Junho}{Myung}
\Author[1]{Kiwoong}{Park}
\Author[3]{Jiheun}{Ha}
\Author[4]{Min-Young}{Choi}
\Author[5, 6]{Bomi}{Kim}
\Author[7]{Hyun W.}{Ka}
\Author[4]{Jung-Hun}{Woo}
\Author[1][alice.oh@kaist.edu]{Alice}{Oh}
\Author[3][hmcjeon@kaist.ac.kr]{Haewon}{McJeon}

\affil[1]{School of Computing, KAIST, Daejeon, Korea}
\affil[2]{Language Technologies Institute, Carnegie Mellon University, Pittsburgh, USA}
\affil[3]{Graduate School of Green Growth and Sustainability, KAIST, Daejeon, Korea}
\affil[4]{Graduate School of Environmental Studies, Seoul National University, Seoul, Korea}
\affil[5]{Environmental Planning Institute, Seoul National University, Seoul, Korea}
\affil[6]{Department of Technology Fusion Engineering, Konkuk University, Seoul, Korea}
\affil[7]{School of Transdisciplinary Studies, KAIST, Daejeon, Korea}




\runningtitle{ML-IAM: Emulating IAMs with ML}

\runningauthor{Shin et al.}

\received{}
\pubdiscuss{} 
\revised{}
\accepted{}
\published{}


\firstpage{1}

\maketitle

\begin{abstract}
Integrated Assessment Models (IAMs) are the standard tools for projecting emissions and energy systems under diverse socioeconomic and technological scenarios. However, their computational intensity limits accessibility and application scope. We present ML-IAM v1.0, to our knowledge the first supervised, input-conditional emulator trained on the IPCC AR6 Scenarios Database. From 42 exogenous scenario inputs and model-family and region identifiers, we train it to predict six primary-energy and three greenhouse-gas emission outputs for 15 IAM families from the global to the country scale. We implement three architectures as a demonstration of the approach, XGBoost, Long Short-Term Memory (LSTM), and the Temporal Fusion Transformer (TFT), which reach a test-set $R^2$ of 0.975, 0.971, and 0.985 against original IAM data. Once trained, ML-IAM v1.0 generates the full test set (up to 866 scenarios) in 14--85\,s, a throughput of 0.1--1\,s per complete multi-region scenario and a speed-up of roughly two to five orders of magnitude over the IAMs themselves. This makes large-ensemble scenario exploration feasible at a cost that is prohibitive for IAMs and positions ML-IAM as a complement to them rather than a replacement.
\end{abstract}


\introduction  
Integrated Assessment Models (IAMs) are computational frameworks that integrate economics, energy systems, land use, and climate science to project how different technological, policy, and socioeconomic scenarios affect future greenhouse gas (GHG) emissions and climate outcomes \citep{IPCC_AR6_WG3_Ch3_2022}. These models are widely used for climate policy analysis, supporting the assessment of mitigation pathways and their economic implications across different policy contexts \citep{RIAHI2017153}. Despite this, IAMs face limitations: they are computationally intensive, with a single scenario taking up to hours to run (for example, GCAM 0.5--8 h, REMIND 3--7 h, MESSAGEix 3 min--4 h, as reported in an expert survey by \citet{li2025using}).

Machine learning (ML) has become an established means of accelerating computationally expensive Earth system simulation, in two distinct forms. The first serves as a data-driven alternative to numerical models: trained on abundant reanalysis and multi-model data \citep{Hersbach2020_ERA5, Eyring2016_CMIP6}, systems such as GraphCast~\citep{lam2023graphcast_science}, ClimaX~\citep{nguyen2023climax}, and Aurora~\citep{bodnar2025aurora} now rival numerical weather and climate forecasts at a fraction of the cost. This form is considerably harder to realize for IAMs. Weather and climate models forecast hours to decades ahead from dense observational records, whereas IAM scenarios project to 2100 under policy and technology conditions well beyond the range of historical experience. The historical emission and energy statistics that do exist \citep{friedlingstein2026gcb} are also far shorter and coarser than the records observation-trained models rely on. The second form, emulation, learns the input--output mapping of a specific simulator from that simulator's own outputs, and has been applied to Earth system models \citep{climatebench, saleem2024pacer} and dynamic global vegetation models \citep{gmd-18-4317-2025}. Deep learning has recently reached IAM scenario data as well, though so far in generative rather than predictive form \citep{li2025using}. ML-IAM therefore takes the emulation route to address the computational limitations described above.

In climate science, the term \textit{emulator} has traditionally referred to reduced-complexity models such as MAGICC~\citep{meinshausen2011magicc}, FaIR~\citep{smith2018fair}, and OSCAR~\citep{gasser2017oscar}. These approximate more complex Earth system models through a small set of simplified physical equations, with parameters calibrated against process-based simulations. They enable rapid scenario exploration at the cost of physical detail. ML-IAM is an emulator in the data-driven sense. Like the ML emulators above, it learns a simulator's input--output mapping directly from the simulator's outputs without imposing explicit equations, but it applies this to IAMs rather than to physical climate variables.

Emulation relaxes the data requirement but does not remove it. It still needs a large body of simulator output, and until recently IAMs could not supply one. The reason lies in how training samples arise. A climate model emulator learns a mapping onto gridded fields. Every grid cell at every output time step is a training sample, so a single run yields millions of them. ClimateBench, for instance, is trained on only a few scenario runs from standardized CMIP experiments \citep{climatebench, Eyring2016_CMIP6}. An IAM emulator instead learns region-level trajectories of a few dozen time steps. Its training set therefore grows only with the number of scenarios. IAM scenarios, however, cannot be produced in bulk. Each is the product of a dedicated research effort, typically a new narrative, policy design, or regional study \citep{peters2023ar6}. Scenarios therefore accumulate slowly and remain scattered across individual studies and model intercomparison projects \citep{guivarch2022using}. They began to be consolidated only in the scenario databases compiled for IPCC assessments \citep{huppmann2018scenario, peters2023ar6}. The 2022 AR6 Scenarios Database, with 3,131 scenarios from 191 modeling frameworks \citep{byers_edward_2022_5886911}, is the first large enough to provide a workable foundation for ML emulation of IAMs.

Reflecting these constraints, data-driven work on IAM output has so far been limited to a handful of studies. None, however, has attempted ML emulation across IAM families (distinct models such as GCAM, REMIND, or MESSAGEix) rather than within a single one. \citet{takakura2020reproducing} developed emulators for the economic impacts of climate change, reproducing the outputs of sectoral biophysical impact models coupled with economic models. These components can be integrated within IAM workflows but do not emulate IAM mitigation pathways themselves. \citet{li2025using} used the AR6 Scenarios Database to train generative models that produce synthetic scenarios for aggregated IPCC climate categories \citep{IPCC_AR6_WG3_Ch3_2022}. Because these scenarios are conditioned only on the target climate category rather than on the socioeconomic and policy inputs that produce it, the approach supports scenario synthesis rather than conditional prediction for policy analysis. \citet{holmes2024emulatingglobalchangeanalysis} developed a deep neural network emulator for GCAM that maps 12~energy sector input factors to 44~outputs spanning energy, land, and water sectors, achieving median $R^2 > 0.99$. This demonstrates the feasibility of high-fidelity multi-sector emulation, but for a single IAM run as a designed sensitivity ensemble \citep{woodard2023scenario} rather than for the heterogeneous, multi-model scenarios of a database. \citet{emIAM2025} take a non-ML route: they fit marginal abatement cost curves to nine IAMs in the ENGAGE database \citep{riahi2021cost} and the GET model and couple them with a simple climate model to reproduce cost-effective emission pathways under climate targets, a framework since extended to 2150 \citep{xiong2025scenario}. emIAM thus achieves systematic cross-model emulation, but only for emission pathways and only conditioned on a climate target rather than on exogenous scenario inputs. Table~\ref{tab:comparison} compares these approaches. None addresses the combined challenge of \textit{predictive} emulation of both energy and emission variables across diverse IAM families from exogenous scenario inputs. A fast, cross-family emulator would enable systematic scenario exploration, such as large-ensemble sensitivity analysis and multi-objective policy screening \citep{lamontagne2018large, woodard2023scenario}, that remains computationally prohibitive with traditional IAMs.

\begin{table*}[bt]
\caption{Comparison of recent IAM emulation approaches. ML-IAM is the only one that predicts both primary energy and emissions from exogenous scenario inputs across many IAM families.}
\label{tab:comparison}
\centering
\footnotesize
\renewcommand{\arraystretch}{1.15}
\begin{tabular}{l l l l l}
\hline
 & \textbf{ML-IAM v1.0} & \textbf{Deep-IAM} & \textbf{GCAMnet} & \textbf{emIAM v1.0} \\[-1pt]
 & (this study) & \citet{li2025using} & \citet{holmes2024emulatingglobalchangeanalysis} & \citet{emIAM2025} \\
\hline
Approach & Predictive emulation & Scenario synthesis & Single-model emulation & MAC curves + \\[-1pt]
 & (supervised ML/DL) & (generative DL) & (supervised DL) & simple climate model \\[5pt]
Database & AR6 full & AR6 vetted & GCAM ensemble & ENGAGE + GET \\[5pt]
IAM families & 15 & 13 & 1 & 10 \\[5pt]
Regions & 59 (country to global) & Global only & 32 GCAM regions & 10 + global \\[5pt]
Time steps & $\le$20 (5-yr, to 2100) & 9 (10-yr, to 2100) & 16 (5-yr, to 2100) & 81 (annual, to 2100) \\[5pt]
Inputs & 42 exogenous (techno- & Climate category & 12 energy sector & Climate target \\[-1pt]
 & economic, macro, policy) & & factors & \\[5pt]
Outputs & 6 primary energy & ${\sim}$6 energy and & 44 energy, land, & 6 emissions \\[-1pt]
 & + 3 emissions & emission variables & and water variables & \\[5pt]
Input-conditional & Yes & No & Yes & Climate target only \\
\hline
\end{tabular}
\end{table*}

\begin{figure}[bt]
    \centering
    \includegraphics[width=0.85\linewidth]{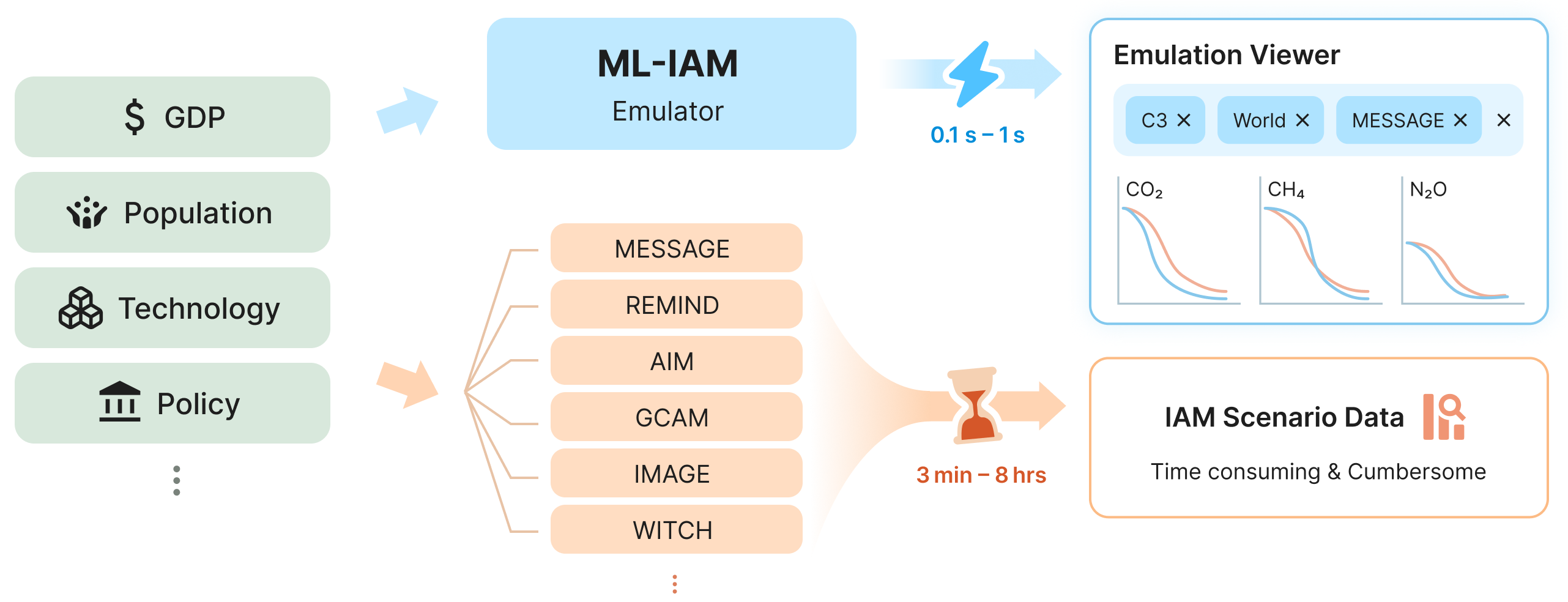}
    \caption{ML-IAM v1.0 emulates multiple IAM families from exogenous scenario inputs (GDP, population, technology costs, policy variables), generating emission and primary energy projections for the test set in 14--85 seconds (866 scenarios with XGBoost and LSTM, 646 with TFT), or 0.1--1 seconds per complete multi-region scenario (Figure~\ref{fig:computational_cost}), compared to 3 minutes to 8 hours per scenario for traditional IAMs. An interactive Emulation Viewer enables rapid scenario exploration across regions and model families.}
    \label{fig:concept}
\end{figure}

We present ML-IAM v1.0, to our knowledge the first supervised, input-conditional emulator trained on the IPCC AR6 Scenarios Database (Figure~\ref{fig:concept}). ML-IAM predicts GHG emissions and primary energy by source from exogenous inputs such as GDP, population, technology costs, and the carbon price, covering 15 IAM families from the global to the country scale. It learns these relationships across families rather than within a single one. We test three architectures: XGBoost \citep{chen2016xgboost}, a gradient-boosted tree ensemble; LSTM \citep{hochreiter1997long}, a recurrent neural network for sequential data; and the Temporal Fusion Transformer \citep[TFT;][]{lim2021temporal}, an attention-based model for multi-horizon forecasting. On held-out regional trajectories they reach an $R^2$ of 0.975, 0.971, and 0.985, respectively. They generate the full test set in 14 to 85 seconds, or 0.1 to 1 second per complete multi-region scenario, roughly two to five orders of magnitude faster than the IAMs themselves. To examine which inputs drive the predictions, we apply SHAP analysis \citep{lundberg2017unified}, which quantifies how much each input variable contributes to a given prediction. ML-IAM thus offers a fast complement to IAMs for large-ensemble scenario exploration, and we release it with an interactive Emulation Viewer for browsing projections across regions and IAM families.

\section{Methods}
\label{sec:methods}
\begin{figure}[tb]
    \centering
    \includegraphics[width=\linewidth]{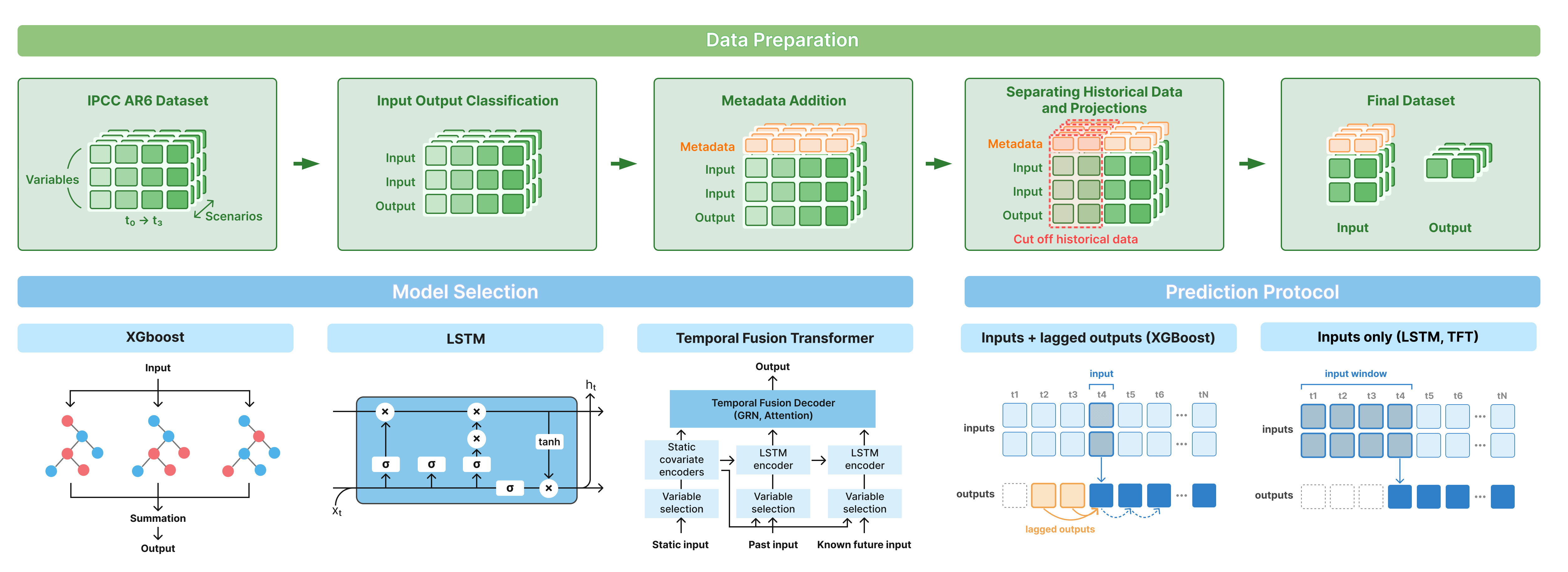}
    \caption{Framework of ML-IAM. \textbf{Data Preparation (top):} The pipeline processes the IPCC AR6 Scenarios Database through three stages: input/output variable classification, metadata addition (model family and region identifiers), and separation of historical data from projections (removing pre-base-year data). \textbf{Model Selection (left):} Three architectures are evaluated: XGBoost for tabular modeling, LSTM for sequential dependencies, and Temporal Fusion Transformer (TFT) for attention-based multi-horizon forecasting. \textbf{Prediction Protocol (right):} All models predict outputs from a window of the scenario's inputs; no model sees outputs beyond the start of a trajectory. XGBoost additionally reads the two preceding outputs, reported at first and thereafter its own predictions (reported throughout training).}
    \label{fig:framework}
\end{figure}

In this section, we outline the overall model training process, including the dataset used for model training (Section \ref{sec:method_dataset}), the machine learning model selection (Section \ref{sec:method_selection}), model training (Section \ref{sec:method_model_setting}), and model interpretability analysis (Section \ref{sec:method_interpretability}). Hereafter, ``model'' refers to machine learning models unless otherwise specified, while ``IAM'' refers to Integrated Assessment Models. Figure \ref{fig:framework} provides an overview of the complete ML-IAM framework.

\subsection{Dataset for Model Training}
\label{sec:method_dataset}
For model training, we use the IPCC AR6 Scenarios Database~\citep{byers_edward_2022_5886911}, which includes 3,131 scenarios from 191 modeling frameworks and over 95 model families, covering global, multi-regional aggregations, and country-level scales. As the largest standardized collection of IAM scenarios to date, its diversity in models, regions, and scenarios enables machine learning models to identify underlying patterns across different IAM frameworks rather than fitting to specific model characteristics. Specifically, inter-model spread, the variation in outputs across different IAM families for the same scenario, is captured through the model family categorical identifier (Section~\ref{sec:method_metadata}), while intra-model spread, the variation across scenarios within a single IAM family, is captured through the exogenous input variables that differ across scenarios. However, variables differ widely across models; therefore, reporting coverage varies dramatically—over half of the variables appear in fewer than 10\% of scenarios. Our data preparation pipeline involves: (1) classification of input and output variables, (2) addition of model family and region identifiers as metadata, and (3) separation of historical data from projections.

We use the database's original, unharmonized outputs. Harmonization~\citep{gidden2019global} is a separate post-processing step that aligns IAM outputs to a common historical reference, primarily for downstream use as CMIP climate model inputs. Because ML-IAM aims to emulate IAM behavior directly, training on unharmonized outputs preserves each family's characteristic response patterns. Harmonization can still be applied to ML-IAM predictions afterward if desired, since the procedure itself is not computationally intensive.

\subsubsection{Input Output Classification}
Unlike typical ML datasets, IPCC AR6 Scenarios Database variables are not pre-labeled as inputs or outputs, as their roles vary by modeling objective. Exogenous variables such as population, GDP, carbon prices, and capital costs consistently function as model inputs across IAMs. In contrast, endogenous variables such as emissions and energy production are typically calculated by IAMs based on these inputs and internal model dynamics. For ML-IAM, we use only exogenous variables as inputs for broader generalizability.

\textit{Exogenous IAM Variables.} Proper classification is critical to prevent data leakage. If variables influenced by target outputs are incorrectly used as inputs (e.g., using coal consumption to predict emissions), models can achieve misleadingly high performance by exploiting shallow dependencies rather than causal mechanisms. We engaged domain experts to classify all 1,624 variables in the database as inputs or outputs. We keep the 534 most frequently reported variables, those reported for at least 10,100 model--scenario--region combinations. The cutoff was set manually and not optimized, so other values could be explored in future work. Of these, 42 are classified as inputs, and we use them.

\textit{Output Variables.} Our implementation focuses on predicting nine key endogenous variables: energy production from six sources (coal, gas, nuclear, oil, solar, wind) and three GHG emissions (CO$_2$, CH$_4$, N$_2$O). However, the framework readily extends to any endogenous variable in the database. We use aggregate variables rather than sectoral ones (e.g., Emissions|CO2|Energy, Final Energy|Industry). Sectoral variables are reported by fewer model families, and their definitions differ between IAMs. Table~\ref{tab:variables_summary} summarizes the input and output variables, with the complete list provided in Appendix Table~\ref{tab:appendix_variables}.

\begin{table}[tb]
\small
\centering
\caption{Summary of ML-IAM input and output variables. The full list of individual variables is provided in Appendix Table~\ref{tab:appendix_variables}.}
\label{tab:variables_summary}
\begin{tabular}{lll}
\hline
\textbf{Role} & \textbf{Category} & \textbf{Count} \\
\hline
\multirow{2}{*}{Metadata}
 & Model Family & -- \\
 & Region & -- \\
\hline
\multirow{7}{*}{Input (42 variables)} 
 & Capital Cost (electricity generation by source) & 14 \\
 & Efficiency (power plant by source) & 6 \\
 & Lifetime (power plant by source) & 4 \\
 & Fixed O\&M Cost (electricity by source) & 13 \\
 & Macroeconomic (GDP|MER, GDP|PPP, Population) & 3 \\
 & Policy and other (Price|Carbon, Yield|Cereal) & 2 \\
\hline
\multirow{2}{*}{Output (9 variables)} 
 & Primary Energy (Coal, Gas, Oil, Nuclear, Solar, Wind) & 6 \\
 & Emissions (CO$_2$, CH$_4$, N$_2$O) & 3 \\
\hline
\end{tabular}
\end{table}

\subsubsection{Metadata Addition}
\label{sec:method_metadata}
Beyond the exogenous IAM variables, we augment each data instance with model family and region identifiers as categorical inputs.

\textit{Model Family.} We explicitly encode model family as a categorical input rather than treating model-specific behaviors as noise. This improves predictive accuracy by capturing each IAM's characteristic response patterns.

\textit{Region.} For computational tractability, we treat regions independently, without modeling inter-regional interactions such as trade flows. This assumption is less restrictive given the dataset composition: 70\% of data points represent aggregated regions (R10 to World level) where inter-regional dynamics are already embedded in the aggregation, while only 30\% are country-level regions. Each region within a scenario thus constitutes a separate data point, allowing us to scale to the full geographic scope of IAM scenarios. To support scale-stratified downstream steps, each region is additionally tagged with a geographic scale label (\textit{Region\_Scale}).

\subsubsection{Separating Historical Data and Projections}
IAMs initialize projections from a base year that separates the historical period (used for calibration and validation with observed data) from the future projection period (generated through model dynamics). Including historical observations in model training would contaminate the learning process by mixing empirical data with model-generated projections. The AR6 Scenarios Database does not record each model's base year, so we identified it from model documentation. To minimize manual effort, we first filter the dataset to remove records with no reported output or with fewer than 40\% of the 51 selected variables reported (i.e.\ fewer than 20). For the remaining data, we retain only the projection period by removing pre-base-year historical observations. For models whose base years are publicly documented, we apply their specified cutoffs. Otherwise, we conservatively use 2020 as the threshold. The base years we collected are included with the source code \citep{yenshin1891_2026_24248338}. All series are truncated at 2100 and kept at their reported time steps (Section~\ref{sec:method_model_setting}).

The final dataset contains 23,462 time series from 1,120 scenarios, each series representing a distinct model-scenario-region combination, providing sufficient scale and diversity for training. Of the more than 95 model families in the database, only 15 (43 model versions) pass these filters. The others fall below the completeness threshold because they do not report the techno-economic cost, efficiency, and lifetime parameters, which make up 37 of the 42 inputs. One further family reports the inputs but none of the nine outputs.

\subsection{Machine Learning Model Selection}
\label{sec:method_selection}
The task combines tabular and time-series characteristics (Section~\ref{sec:method_task_char}). Rather than searching over architectures, we pose it in three formulations that differ in how a window of scenario inputs is turned into outputs (Section~\ref{sec:method_protocol}) and train one widely used model in each: XGBoost \citep{chen2016xgboost} as a recursive one-step tabular predictor, LSTM \citep{hochreiter1997long} as a direct one-step regressor over an input window, and the Temporal Fusion Transformer (TFT) \citep{lim2021temporal} as a direct multi-horizon forecaster. The comparison among them shows that the task can be solved in each form. It is not a ranking of architectures, and other members of each formulation may do as well or better. Appendix~\ref{appendix:model_review} explains which architectures were left out and why.

\subsubsection{Task Characteristics}
\label{sec:method_task_char}
IAMs produce trajectories of system states at fixed intervals (typically 5 years) from a model-specific base year through 2100, in up to 20 steps with a median of 16. How successive states are linked depends on the solution concept: recursive-dynamic models (e.g.\ GCAM, IMAGE) compute each timestep from the previous state and the current assumptions, whereas intertemporal-optimization models (e.g.\ REMIND, WITCH) solve the entire horizon simultaneously with perfect foresight of future assumptions. In either case, two properties set IAM emulation apart. Unlike standard time-series forecasting, which extrapolates from a series' own past, a trajectory is driven mainly by the scenario's assumptions. Unlike tabular regression, which predicts a target from a fixed set of features at a single point in time~\citep{borisov2022deep}, adjacent states are tightly coupled through physical, economic, and technological relationships. Three characteristics of the task guided our choice of models:

\noindent\textbf{Short, non-periodic series:} Series contain at most 20 timesteps (median 16) and have no seasonal pattern. Adjacent states are strongly coupled, but dependence over long spans is weak, so models designed for long context windows (50--100+ timesteps) gain nothing here and in some cases cannot be applied.

\noindent\textbf{Covariate-driven:} Outputs depend on the interaction of over 40 input variables and, under perfect foresight, on their future as well as current values.

\noindent\textbf{Many heterogeneous series:} One model must cover diverse combinations of model families, regions, and scenarios, and predict several outputs for each.

Consequently, the three formulations of Section~\ref{sec:method_protocol} each use a method that can exploit covariates: a tabular learner applied recursively, and sequence models built for short input windows.

\subsubsection{Related Work and Applicability}
IAM emulation is closely related to forecasting in economics, where annual or quarterly series are short and forecasts depend on many input variables rather than on repeating temporal patterns \citep{yoon2021forecasting, yang2024machine}. In that domain, tree-based models, recurrent neural networks, and attention-based architectures have each been used with success \citep{yoon2021forecasting, anesti2024forecasting, wang2021economic, dong2025gross, han2023economic}, and gradient-boosted trees remain the strongest general-purpose choice for heterogeneous tabular data \citep{shwartz2022tabular, grinsztajn2022why, borisov2022deep}. Most recent time-series architectures, by contrast, are built for the opposite regime: hundreds of timesteps with seasonal structure and little covariate information, against at most 20 policy-driven steps here. Gaussian-process emulators, widely used in the geosciences \citep{climatebench}, are suited to training sets of a few hundred points rather than the hundreds of thousands of rows available here. Appendix~\ref{appendix:model_review} explains why these architectures were not considered.

\subsubsection{Selected Models}
We train one representative of each formulation:

\textbf{XGBoost}~\citep{chen2016xgboost} is a tree-based model that builds many decision trees in sequence, each correcting the errors of the previous ones. It excels at combining many input variables to make predictions, is used in economic forecasting \citep{yoon2021forecasting}, and is the representative of the tabular formulation because gradient-boosted trees match or outperform neural tabular models across benchmarks \citep{shwartz2022tabular, grinsztajn2022why}. Because it treats each timestep independently, it relies on lag features (values from previous timesteps, such as the previous timestep's emissions) to incorporate temporal information.

\textbf{LSTM} (Long Short-Term Memory)~\citep{hochreiter1997long} is a neural network designed for sequential data. It processes the inputs of consecutive timesteps in order, carrying a hidden state from one to the next, which suits IAMs' step-by-step projection structure where each year depends on the previous state plus current input assumptions.

\textbf{TFT} (Temporal Fusion Transformer)~\citep{lim2021temporal} is a neural network that uses an attention mechanism---a method for dynamically identifying which inputs are most relevant at each prediction step---to forecast multiple timesteps ahead simultaneously. It is designed to handle diverse input types (e.g., static metadata like model family alongside time-varying inputs like carbon price), which matches the structure of this task directly. Architectural details for all three models are provided in Appendix~\ref{appendix:training_details}.

These three do not exhaust their formulations. Other gradient-boosting implementations (LightGBM, \citealp{ke2017lightgbm}; CatBoost, \citealp{prokhorenkova2018catboost}) or a multilayer perceptron for the tabular formulation, gated recurrent units \citep{cho2014learning} for the windowed regressor, and other sequence-to-sequence forecasters \citep{sutskever2014sequence} for the multi-horizon case are natural next comparisons that we did not run. We chose to apply one tuning protocol identically to three formulations rather than to add architectures tuned under different budgets (Appendix~\ref{appendix:hpo}), and we release the train/validation/test assignment so that further models can be scored on the same series.

As a reference for these three, we also fit a linear baseline: an untuned ridge regression per output on the same features and splits as XGBoost, and evaluated with the same recursive rollout (Section~\ref{sec:method_protocol}). We use the ridge penalty because ordinary least squares diverges in the rollout (Appendix~\ref{appendix:model_settings}). The gap between this baseline and each emulator measures what nonlinear modeling adds.

\subsection{Model Training}
\label{sec:method_model_setting}
This section describes the training pipeline, covering data splitting, preprocessing, the training and evaluation protocol, and hyperparameter optimization.

\subsubsection{Data Splitting}
\label{sec:method_data_splitting}
A scenario is a (Model, Scenario) pair: one IAM's run of one scenario, covering all of the regions that model reports, 21 on average in our dataset. Each region of a scenario forms one series of projected values from the base year through 2100. The dataset holds 1,120 scenarios and 23,462 regional series, and we split it at the series level: series are divided into training (80\%), validation (10\%), and test (10\%) sets by a group-based split with a fixed random seed, so that all timesteps of a series fall in one partition, and the test set comprises 2,346 series from 866 scenarios. This prevents temporal leakage, where a model could exploit future observations of a series already present in the training set.

The split is at the series level, so test performance measures how well the emulator reproduces a held-out regional trajectory of a scenario whose model family, region, and protocol it has seen. Scenario categories (C1--C8) appear in every split, as does every region and model family with at least 20 series (Appendix~\ref{appendix:model_settings}). Most test scenarios are also partly seen during training. For 98\% of them, another region of the same scenario is in the training or validation set, and for 80\% the World aggregate is. Many AR6 scenarios also come from multi-model protocols (e.g. ENGAGE, NGFS, the SSPs), in which several IAMs ran the same scenario name from near-identical exogenous inputs, so 57\% of test scenarios have a same-name scenario from a different model in the training or validation set. Appendix~\ref{appendix:counterparts} tests whether these counterparts account for the scores (Table~\ref{tab:counterparts}). A scenario's World aggregate makes no difference, and a same-name run from another model matters little for any of the three models (a gap in $R^2$ of at most 0.015). The scores therefore reflect what the emulators learn from a scenario's inputs rather than from its siblings. Generalization to entirely new scenarios, or to model families and regions absent from training, is a different and harder test that would require broader data coverage, and we leave it for future work. Likewise, ML-IAM has not been validated for input values outside the ranges observed in the training data, such as extreme GDP trajectories or policy assumptions absent from the AR6 database.

\subsubsection{Preprocessing}
\textit{Lag Features.}
XGBoost receives the output values of the preceding timesteps as input features (hereafter, lag features), e.g.\ the previous values of CO$_2$ emissions and coal production. Because the model treats each timestep as an independent tabular prediction, lag features are its only mechanism for incorporating temporal information. LSTM and TFT receive no lag features: they consume the inputs of several consecutive timesteps directly.

\textit{Missing Data Handling.}
Output values that an IAM did not report are excluded from the training loss and from every metric through a per-element observation mask, so that a series with a partially reported output set is still used for the outputs it reports; rows in which no output is reported are dropped. For XGBoost, whose lag features require complete output vectors, a gap in an output series is filled for lag use only with the last reported value before it. For input features, the treatment differs across architectures. XGBoost natively handles missing values through its sparsity-aware split finding algorithm, allowing preservation of the original data structure without imputation. In contrast, LSTM and TFT required explicit missing data handling through a two-part strategy: (1) introducing binary missingness indicators for each input variable, and (2) imputing missing values with the training-set median of the same variable and regional scale. This approach enables the models to distinguish between actual observations and imputed estimates while enabling gradient-based training. However, median-filled values may still introduce residual noise relative to XGBoost's native sparsity handling.

Reporting intervals vary across series: 36\,\% report at five-year steps throughout, 53\,\% switch from five- to ten-year steps later in the century, and 10\,\% report at ten-year or coarser steps throughout. All three models treat the reported timesteps of a series as consecutive steps regardless of their spacing, and none receives the year as an input. All input and output variables are standardized using z-score normalization fitted on the training set, ensuring comparable scales across outputs measured in different units (e.g., PJ/yr vs. Mt CO$_2$/yr).

\subsubsection{Prediction Formulations and Evaluation Protocol}
\label{sec:method_protocol}
All three models share one design principle: they generate a scenario's outputs from a window of that scenario's \textit{inputs}, not from observed outputs, so that the emulator can be run for a scenario that no IAM has executed. No model is given a scenario's outputs beyond the start of its trajectory: only XGBoost is seeded with reported outputs, and only those of the two timesteps preceding its first prediction. The models differ in how the input window is used. The length of the window, the context length $L$, is treated as a hyperparameter and searched over the same values ($L \in \{2, 3\}$) for all three models (Appendix~\ref{appendix:hpo}); the selected values are $L = 2$ for XGBoost and $L = 3$ for LSTM and TFT.

\textbf{XGBoost} is a \textit{recursive one-step} predictor with one booster per output: each booster predicts its output at a timestep from the current inputs and that output's values at the $L$ preceding timesteps. During training the lagged outputs are the reported values (teacher forcing). At evaluation the model is rolled out along each trajectory, seeded with the reported values of the two preceding timesteps and thereafter conditioning on its own predictions, so that errors can compound within an output. XGBoost is evaluated on the rollout throughout, including during hyperparameter search.

\textbf{LSTM} is a \textit{direct one-step} regressor: it predicts the outputs of a timestep from the inputs of that timestep and the $L-1$ preceding ones, with learned embeddings for the \textit{Model\_Family} and \textit{Region} identifiers. It receives no output values, so each timestep is predicted independently and no error propagates between timesteps.

\textbf{TFT} is a \textit{direct multi-horizon} forecaster: from the inputs of the $L$ preceding timesteps (encoder) and the twelve timesteps to be predicted (decoder), it produces all twelve outputs in a single pass, with the \textit{Model\_Family} and \textit{Region} identifiers as static categoricals with learned embeddings. Its attention is causal, so the prediction for a timestep draws on the inputs up to and including that timestep, not on those of later timesteps. Like the LSTM it receives no output values. Because a single pass covers $3+12=15$ timesteps and test trajectories have up to 20, each trajectory is evaluated with two windows: an early window over its first 15 timesteps, predicting timesteps 4--15, and a late window over its last 15, predicting the final 12. Where the two overlap, their predictions are combined with a linear weight that shifts from the early to the late window across the overlap. The first three timesteps of a trajectory receive no prediction, and trajectories shorter than 15 timesteps are excluded from the TFT evaluation. Implementation details for the three formulations are provided in Appendix~\ref{appendix:prediction_protocols}.

Hyperparameters are selected by a two-stage random search with the same protocol for all three models (Appendix~\ref{appendix:hpo}), scored on the validation set under each model's evaluation formulation. Each model uses a single hyperparameter configuration for all nine outputs (Appendix~\ref{appendix:model_settings}). The LSTM and the TFT predict the nine outputs with one network; XGBoost fits one booster per output. The selected configuration is trained on the training set with early stopping on the validation set, and the resulting model is evaluated on the held-out test set. We report $R^2$ as the primary metric, computed per output over observed test elements and averaged across the nine outputs; RMSE and MAE, and the same metrics on the training and validation sets, are reported in Appendix~\ref{appendix:extended_results}.

\subsection{Model Interpretability Analysis}
\label{sec:method_interpretability}
We use SHAP (SHapley Additive exPlanations; \citet{lundberg2017unified}) to examine which inputs drive each model's predictions. SHAP assigns each feature a contribution to an individual prediction such that the contributions sum to the prediction's deviation from the average. We apply TreeExplainer for XGBoost and expected gradients \citep{erion2021improving} for LSTM and TFT (Appendix~\ref{appendix:shap_details}). Attributions are computed on a sample of test scenarios at the R10 scale and, for the sequence models, summed over the input window. The resulting beeswarm plots rank features by mean absolute attribution and color each point by the feature's value, showing the direction as well as the size of its effect. Lag features, which only XGBoost uses, are reported as they are; a large attribution to a lagged output therefore reflects the recursive formulation and not an exogenous driver. SHAP values indicate associations, not causal relationships. For the LSTM and TFT they can also vary with the choice of background samples. The patterns reported below should therefore be read as suggestive.

\section{Results}
We evaluate the three emulators on the held-out test set (Section~\ref{sec:results_eval}) and examine what they rely on through SHAP analysis (Section~\ref{sec:results_interp}). Metrics are computed only over the output values that the IAMs reported. Appendix~\ref{appendix:extended_results} gives the same metrics for the training and validation sets.

\subsection{Performance Evaluation}
\label{sec:results_eval}

\subsubsection{Overall accuracy}
\label{sec:results_overall}
\begin{figure}[htb]
\centering
\includegraphics[width=\linewidth]{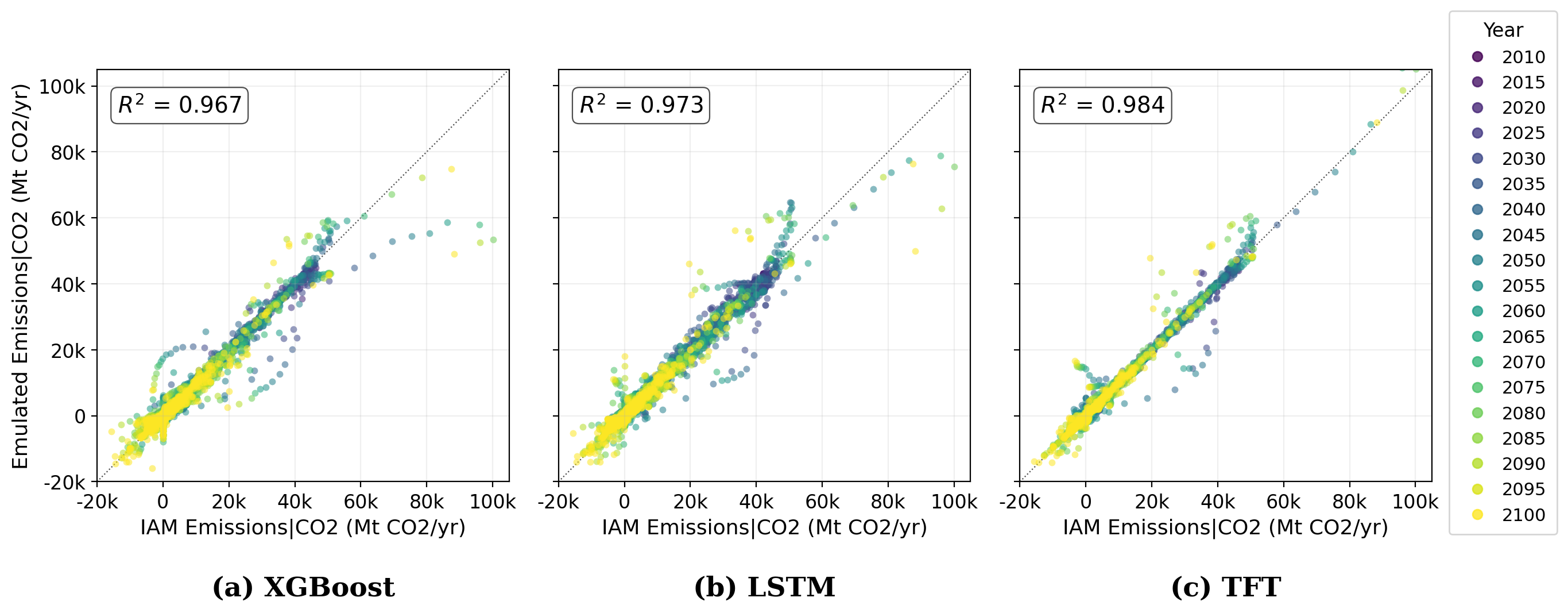}
\caption{ML model predictions versus original IAM CO$_2$ emissions for (a) XGBoost, (b) LSTM, and (c) TFT. Each point represents a test set prediction, colored by projection year.}
\label{fig:scatter_CO2_all}
\end{figure}

TFT is the most accurate emulator (per-target average $R^2 = 0.985$), followed by XGBoost (0.975) and LSTM (0.971) (Table~\ref{tab:r2_by_family}). The linear baseline (ridge regression on XGBoost's features and rollout; Appendix~\ref{appendix:model_settings}) reaches 0.710 (Appendix Table~\ref{tab:performance_summary}), well below all three ML models. This score still flatters a linear emulator: on the C3 World trajectories of Fig.~S5 it reaches $R^2 = 0.31$ against XGBoost's 0.93 (Figs.~S4 and S8; Appendix~\ref{appendix:model_settings}). Figure~\ref{fig:scatter_CO2_all} illustrates the fit for CO$_2$ emissions ($R^2$: 0.967, 0.973, 0.984). XGBoost's and LSTM's largest departures are at the highest emission levels, where both underpredict the few World-scale series above 60\,Gt. TFT tracks them. The other outputs behave alike (Figs.~S1--S3).

\begin{figure}[htb]
\centering
\includegraphics[width=0.9\linewidth]{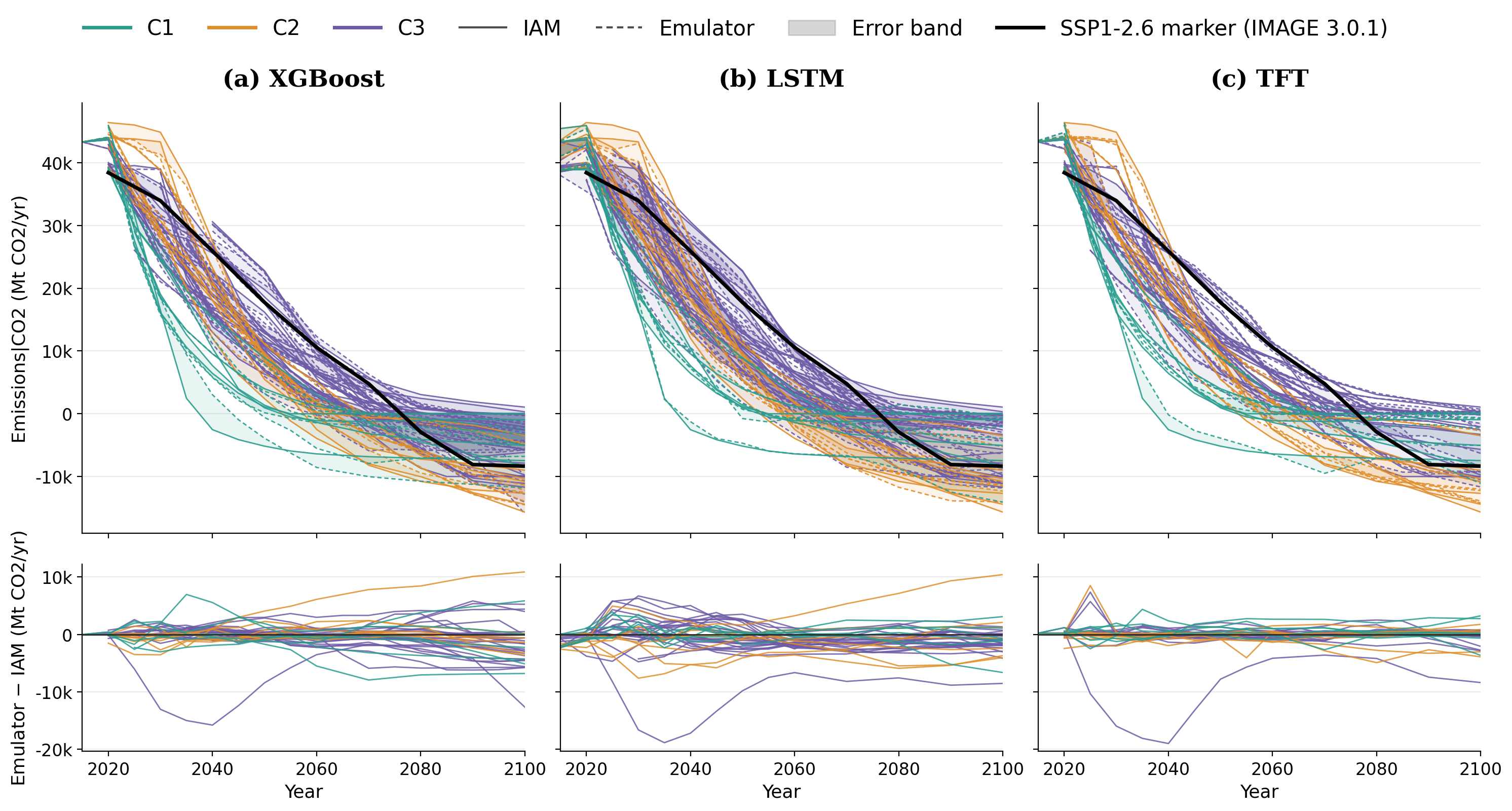}
\caption{CO$_2$ emission trajectories of test-set World trajectories, by climate category (C1: limit warming to 1.5\textdegree C with no or limited overshoot; C2: return to 1.5\textdegree C after a high overshoot; C3: limit warming to 2\textdegree C, $>$67\%), for (a) XGBoost, (b) LSTM, and (c) TFT. Top: IAM projections (solid), emulator predictions (dashed, same color), and their absolute difference (shaded), from each IAM's base year. The thick black line is the SSP1-2.6 marker scenario (IMAGE 3.0.1), shown as a reference; it is not part of the training or test data. Bottom: emulator minus IAM.}
\label{fig:trajectory_CO2_all}
\end{figure}

Beyond aggregate accuracy metrics, the trajectories of individual scenarios show how well each emulator captures the dynamics of emission pathways. Figure~\ref{fig:trajectory_CO2_all} shows CO$_2$ trajectories for the C1--C3 test scenarios in the World region. All three emulators reproduce the shape of each category, including C1's steep near-term decline and C3's mid-century decline to net zero (other outputs in Figs.~S5--S7). The error panels show two clear misses: all three models underpredict one C3 scenario around 2030 to 2040, and XGBoost and LSTM overpredict one C2 scenario by a growing margin after 2060. An interactive Emulation Viewer (\url{https://mliam.dev/}; Appendix~\ref{appendix:viewer_section}) allows these results to be explored by scenario, region, and model family. Its what-if view runs the trained emulators live: a user starts from the inputs of an AR6 scenario, moves carbon price, technology costs, population, or GDP within the training range, and obtains the emulated trajectories (Appendix Fig.~\ref{appendix:viewer_whatif}).

\subsubsection{Breakdown by output, regional scale, and IAM family}
\label{sec:results_breakdown}
By output, TFT is the most uniform, with per-target $R^2$ spanning 0.976--0.992, against 0.941--0.983 for LSTM and 0.951--0.992 for XGBoost (Appendix Table~\ref{tab:r2_results}). By regional scale, all three models are weakest at the World level (XGBoost 0.938, LSTM 0.919, and TFT 0.961) and reach at least 0.95 at every scale from R5 to country level (Appendix Table~\ref{tab:performance_summary}).

\begin{table}[tb]
\small
\centering
\caption{Test-set $R^2$ (per-target average) by IAM family. $n$ is the number of test trajectories scored; XGBoost and LSTM are scored on the same 2,346 trajectories, TFT on the 1,738 that are at least 15 timesteps long. $R^2$ within a family is computed against that family's own variance, so families with little variance on a target are penalized heavily for modest absolute errors. Three families with three or fewer test trajectories are omitted from the rows but included in the totals.}
\label{tab:r2_by_family}
\begin{tabular}{lrrrrr}
\tophline
\textbf{Family} & $n$ (XGB, LSTM) & \textbf{XGBoost} & \textbf{LSTM} & $n$ (TFT) & \textbf{TFT} \\
\middlehline
REMIND    & 602 & 0.975 & 0.969 & 593 & 0.987 \\
MESSAGEix & 417 & 0.981 & 0.978 & 363 & 0.991 \\
WITCH     & 236 & 0.962 & 0.965 & 236 & 0.969 \\
IMAGE     & 223 & 0.959 & 0.958 & 218 & 0.983 \\
TIAM      & 205 & 0.972 & 0.948 & --  & --    \\
IMACLIM   & 197 & 0.794 & 0.861 & 99 & 0.883 \\
POLES     & 163 & 0.995 & 0.991 & --  & --    \\
GEM       & 139 & 0.990 & 0.988 & 100 & 0.998 \\
AIM       & 110 & 0.992 & 0.989 &  92 & 0.991 \\
GCAM      &  48 & 0.812 & 0.800 &  37 & 0.754 \\
\middlehline
\textbf{All} & 2{,}346 & 0.975 & 0.971 & 1{,}738 & 0.985 \\
\bottomhline
\end{tabular}
\end{table}

By IAM family, accuracy is also even (Table~\ref{tab:r2_by_family}), with one notable exception: GCAM is among the two hardest families for all three models (0.81, 0.80, and 0.75), and the reason is the data rather than the models. AR6 contains 186 GCAM scenarios across 11 model versions, or 3,646 model--scenario--region trajectories, but the processed dataset retains 24 scenarios and 470 trajectories (13\,\%). The completeness filter (Section~\ref{sec:method_dataset}) keeps a row only if it reports at least 20 of the 51 selected variables. The 24 retained scenarios are the only ones that report capital costs, whose 12 variables lift them from about 10 reported variables to just over this threshold. Efficiency, lifetime, and fixed O\&M cost are not reported in any GCAM version. GCAM therefore enters training with the smallest sample of any large family and three of its four techno-economic input groups imputed and flagged as missing on every row, so its outputs are predicted largely from macroeconomic drivers, the carbon price, and the model family embedding. More complete reporting of model inputs could improve these scores (Section~\ref{sec:discussion_reporting}). The other weak family is IMACLIM (0.79, 0.86, and 0.88). It reports CH$_4$ only at the World level, so that score rests on at most seven test trajectories and pulls its average down.

The models are not scored on identical sets (the $n$ columns of Table~\ref{tab:r2_by_family}). Each is scored on the timesteps it predicts (Section~\ref{sec:method_protocol}): LSTM skips the first two timesteps of a trajectory, and XGBoost and TFT the first three. TFT also cannot score the 608 trajectories shorter than its 15-step window, including all of POLES and TIAM. However, TFT's lead does not come from this smaller set. POLES and TIAM are among the easiest families for XGBoost and LSTM, and TFT scores highest in every family it scores except AIM (0.991 against XGBoost's 0.992) and GCAM.

\subsubsection{Computational cost}
\label{sec:results_cost}

\begin{figure}[tb]
    \centering
    \includegraphics[width=0.8\linewidth]{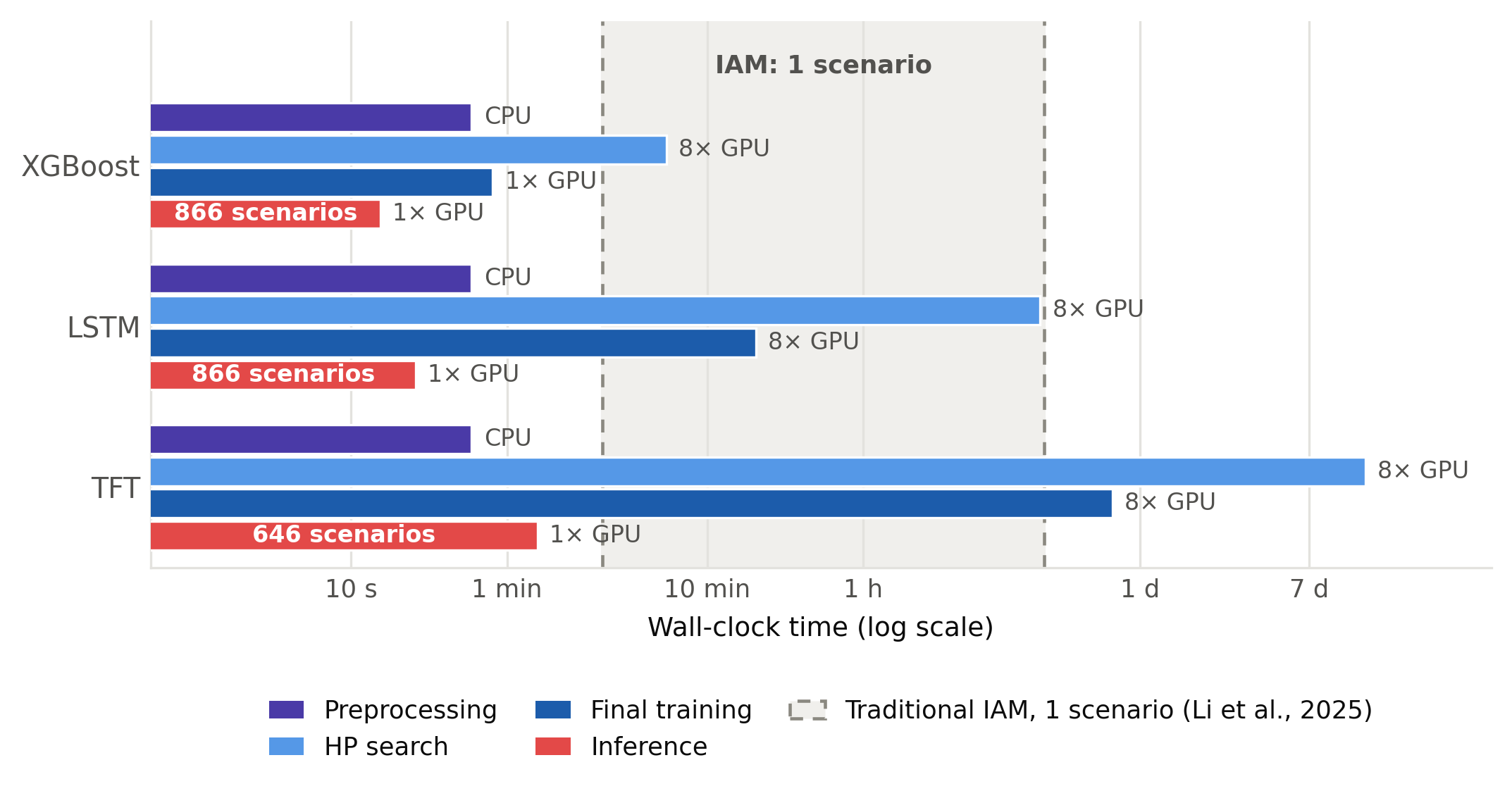}
    \caption{Computational cost by pipeline phase (log-scale time axis). The shaded band marks the 3\,min--8\,h per-scenario runtime of a traditional IAM~\citep{li2025using}. Preprocessing (40\,s, CPU) is shared by all models. All other phases ran on NVIDIA RTX 2080~Ti GPUs, and labels give the GPU count per phase. The TFT search was distributed over three machines (eight, seven, and eight GPUs). Its bar is the single-machine equivalent, the three stage-1 shards' active wall time summed plus the stage-2 refit. Inference times cover the test set end-to-end: 866 scenarios (2,346 regional series) for XGBoost and LSTM, and the 646 scenarios (1,738 regional series) long enough for TFT's 15-step encoder window.}
    \label{fig:computational_cost}
\end{figure}
 
One-off costs differ by orders of magnitude (Fig.~\ref{fig:computational_cost}). Hyperparameter search took 6.1\,min for XGBoost, 7.6\,h for LSTM, and 320\,h for TFT (single-machine equivalent; 185\,h of wall-clock time with stage~1 spread over three machines), and final training 51\,s, 17.5\,min, and 17.6\,h. Once trained, all three generate the test set in under two minutes on one GPU: 14\,s and 21\,s for its 866 scenarios (2,346 regional series) with XGBoost and LSTM, and 85\,s for the 646 scenarios (1,738 regional series) long enough for TFT's encoder window. This is a batch throughput of 6, 9, and 49\,ms per regional series, including data preparation and model loading. A complete scenario covers 21 regions on average, so this amounts to 0.1--1.0\,s per scenario. The latency of a single query is higher: on a CPU, loading a model takes 4--11\,s once, and each series then takes 0.03\,s (LSTM), 0.1\,s (XGBoost), or 0.9\,s (TFT). An IAM needs 3\,min--8\,h per scenario~\citep{li2025using}, so the throughput gain is roughly two to five orders of magnitude per scenario. The two are not strictly comparable, however, because an IAM run produces hundreds of output variables and the emulators nine.

\subsection{Model Interpretability}
\label{sec:results_interp}
\begin{figure}[bt]
\centering
\includegraphics[width=\linewidth]{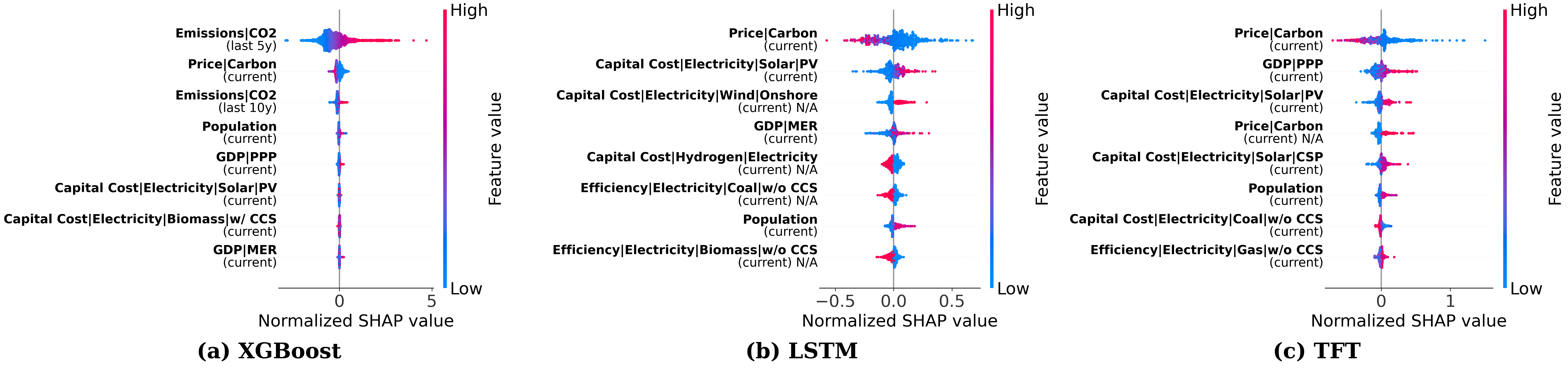}
\caption{Feature importance analysis using SHAP values for CO$_2$ emissions prediction across (a) XGBoost, (b) LSTM, and (c) TFT models. Features are ranked by importance from top to bottom. The horizontal position shows impact: points on the right push emissions higher, points on the left push emissions lower. Colors show the feature value itself: red/pink indicates high values, blue indicates low values. Each point is one sample. The horizontal spread of a row shows how much the feature's effect differs across samples, and the thickness of the swarm shows how many samples have a similar effect. ``Current'' refers to the prediction year; ``last 5y'' and ``last 10y'' refer to the two previous reported timesteps (5 and 10 years earlier at five-year steps). SHAP values are normalized by the sum over features of the output's mean absolute SHAP value.}
\label{fig:shap_CO2_all}
\end{figure}


Figure~\ref{fig:shap_CO2_all} shows SHAP feature importance for CO$_2$ emissions. Figures~S9--S11 show the same for all nine outputs. For XGBoost, whose top feature is necessarily the lagged target, the leading exogenous driver is the carbon price, followed by population and GDP. LSTM and TFT, which see no lagged outputs, both place the carbon price first, with high prices pushing emissions down. Both also rank GDP, the capital cost of solar PV, and population among their top eight features. Across the nine outputs (Figs.~S9--S11), the carbon price is the leading exogenous driver of CO$_2$, coal, and oil in all three models, and GDP or population leads for solar and wind. For gas, nuclear, CH$_4$, and N$_2$O, the leading driver differs by model among the carbon price, GDP, and population.
 
In Fig.~\ref{fig:shap_CO2_all}, the features marked ``N/A'' are missingness indicators (Section~\ref{sec:method_model_setting}), which rank among the top features of the LSTM and, to a lesser extent, the TFT. Reporting coverage varies mainly by model family (Appendix Table~\ref{tab:family_coverage}), so the indicators tell the models where a series comes from. Part of the reporting gap is structural (Section~\ref{sec:discussion_reporting}), so they may also reflect how an IAM represents a technology. Either use is valid for an IAM emulator, but a high rank for an indicator does not mean that the underlying variable is an important driver. XGBoost has no indicator features, because it handles missing inputs natively.

\section{Discussion}

ML-IAM reproduces AR6 IAM output accurately within the range of the database, with all three models reaching an $R^2$ of at least 0.97 and TFT 0.985. We present these results as a demonstration that IAM behavior can be learned from a scenario database with standard ML and DL architectures, not as a ranking of those architectures. The three models solve the task in different forms (Section~\ref{sec:method_protocol}), other designs may do as well or better, and the ranking may shift with the size and reporting format of future databases, such as that of the IPCC Seventh Assessment Report (AR7).

\subsection{Applications}
\label{sec:discussion_applications}

The speedup from hours to about a second per scenario changes what can be done with IAM output within the coverage of a scenario database. Large-ensemble exploration at a scale that is computationally prohibitive with traditional IAMs \citep{lamontagne2018large, woodard2023scenario} becomes routine: controlled experiments that vary one input while holding the others fixed, sensitivity analysis across thousands of scenario variations, uncertainty propagation through Monte Carlo sampling of the inputs, and interactive policy exploration with immediate feedback. The Emulation Viewer's what-if view (Section~\ref{sec:results_overall}; Appendix~\ref{appendix:viewer_section}) is an example of such interactive exploration for one scenario at a time. Run in batch, the same emulators allow input assumptions such as carbon price trajectories, technology costs, and GDP pathways to be varied systematically across the training range. The emulated outputs for each combination can then be screened against policy-relevant criteria, such as the C1--C8 climate categories or sustainable development indicators. This can turn IAM analysis from a small number of carefully chosen runs into a systematic sweep of the input space.

On-demand emulation also serves users outside the IAM community. The AR6 Scenarios Database underpins the IPCC WGIII assessment and is widely used in adjacent fields such as land use, water resources, biodiversity, and air pollution, as well as in policy analysis~\citep{byers_edward_2022_5886911, peters2023ar6}. These users currently select approximate matches from existing databases. With ML-IAM, they can instead generate scenarios tailored to their analytical needs within the range of the database, which makes IAM-emulated projections more customizable than the database they are trained on.

\subsection{Limitations and future work}
\label{sec:discussion_limitations}

Three limitations affect how the results should be read. The first two restrict the evidence to the coverage of the database. First, the data split is by regional trajectory, not by scenario, so a test scenario's other regions can be in training (Section~\ref{sec:method_data_splitting}). The accuracy reported is therefore mostly for scenarios the database already contains; holding out whole scenarios is the next test. Second, ML-IAM has not been validated for input values outside the training distribution (e.g., extreme GDP trajectories, policy assumptions absent from the AR6 database). Predictions in such regimes should be treated with caution. The third is a matter of scale: all three models are weakest at the World level (Table~\ref{tab:performance_summary}), where outputs are far larger than at the country level and are not normalized by regional scale. Predicting trajectories relative to their base year is one way to do so that we have not explored.

Other limitations are matters of scope and reflect choices made for this first version rather than optimized design decisions. Regions are treated independently, without inter-regional interactions such as trade flows and grid connections. Incorporating regional connectivity through architectures like graph neural networks is a natural extension \citep{kipf2016semi}. The input variable set is fixed, so changing it requires retraining, which limits adaptability when data availability varies across models or regions. The nine outputs were chosen for their reporting coverage and do not form closed physical relationships such as an energy balance, so physical-consistency checks were outside this paper's scope. As extensions, the framework accepts further output variables, so Physics-Informed Neural Networks (PINNs) could embed known physical or economic laws directly into training \citep{raissi2019physics, cuomo2022scientific, alonso2023physics, rani2025g}. Finally, ML-IAM returns point predictions without uncertainty estimates. Training an ensemble of models is one way to add them.

\subsection{Implications for scenario reporting}
\label{sec:discussion_reporting}

The coverage of ML-IAM is limited by what the database reports, not by the models. Of the 75 model families that our filters exclude, 74 report none of our techno-economic inputs (Section~\ref{sec:method_dataset}). Part of this gap is structural, because models represent technologies differently and some do not have these parameters \citep{krey2019looking, giarola2021challenges}. Part of it is a matter of reporting. Capital costs alone would bring 42 of the excluded families and 1,410 of their scenarios over our completeness threshold.

A single common format for all inputs may be impractical, because input sets differ across model families. Two smaller steps would still help. The first is to report the inputs that most models share, including at least one group of techno-economic parameters where a model has them. The second is to make model-specific inputs publicly available, for example in code repositories, as several open-source models already do \citep{calvin2019gcam, huppmann2019messageix}. Full disclosure can be difficult for models that are not open source or that rely on licensed data, but we encourage it wherever feasible. Because the number of scenarios grows slowly, fuller reporting of existing ones is a direct way to enlarge the data available for data-driven research on IAMs.

\conclusions  
\label{sec:conclusions}

ML-IAM v1.0 is, to our knowledge, the first supervised, input-conditional emulator of IAMs trained on the AR6 Scenarios Database. XGBoost, LSTM, and TFT reproduce six primary-energy and three emission outputs of 15 IAM families with a test-set $R^2$ of at least 0.97, two to five orders of magnitude faster than the IAMs they emulate. The trained models and an interactive viewer are released with the code. These results come from three architectures, each in one formulation of the task, trained on one dataset; they show that emulation is feasible, not which architecture suits it best. The next steps are to validate on held-out scenarios, to add uncertainty estimates, and, on the data side, to report input assumptions alongside outputs. ML-IAM is a first step toward machine learning that complements conventional IAMs, and it shows what collaboration between modelers, data providers, and machine learning practitioners can produce for climate policy research.




\codedataavailability{
The source code and supporting data files (base year mappings and 
input/output variable classifications) for ML-IAM are permanently 
archived on Zenodo at \url{https://doi.org/10.5281/zenodo.24248338} 
\citep{yenshin1891_2026_24248338}. The code is also available on GitHub at 
\url{https://github.com/YenShin1891/ml-iam}; the archived version is release v1.1.0
(commit \texttt{de82796}). The trained XGBoost, LSTM, and TFT
models reported here are archived on Zenodo at
\url{https://doi.org/10.5281/zenodo.24247480} \citep{shin_2026_24247480}. The
archives hold the model weights, scalers, category vocabularies, and the
train/validation/test assignment of scenario identifiers, and contain no AR6 data.
With the repository, they can be run on new scenarios supplied in the IAMC format. The split
assignment lets a reader with their own AR6 download rebuild the test
set used here. The trained emulators can also be explored
without the training pipeline through the Emulation Viewer at \url{https://mliam.dev/}
(Appendix~\ref{appendix:viewer_section}).
The IPCC AR6 Scenario Database \citep{byers_edward_2022_5886911} is 
managed by IIASA and access instructions are described in 
\url{https://doi.org/10.5281/zenodo.7197970}.
} 



\appendix
\clearpage
\section{Dataset Composition}
\label{appendix:dataset}
Table~\ref{tab:family_coverage} lists the 15 IAM families of the training dataset: how many of their model versions and scenarios the AR6 database holds, how many survive the filters of Section~\ref{sec:method_dataset}, the base-year cutoff applied, and which of the four techno-economic input groups they report. Table~\ref{tab:appendix_variables} lists the 42 inputs and nine outputs with their units and the share of retained series that report each. Preprocessing is described in Sections~\ref{sec:method_dataset} and \ref{sec:method_model_setting} and Appendix~\ref{appendix:model_settings}.

\begin{table}[h]
\footnotesize
\centering
\setlength{\tabcolsep}{4pt}
\renewcommand{\arraystretch}{1.1}
\caption{IAM families of the training dataset. \emph{Versions}: model versions retained / in the AR6 database (versions are assigned to a family by name). \emph{Scenarios}: (model, scenario) pairs in AR6 and retained after filtering; \emph{Series}: retained (model, scenario, region) series. \emph{Base year}: documented base year of the retained versions, below which reported years are removed; an asterisk marks families with a version whose base year is undocumented, for which the cutoff is 2020. The last four columns give the share of retained series that report at least one variable of each techno-economic input group (capital cost, efficiency, lifetime, fixed O\&M cost).}
\label{tab:family_coverage}
\begin{tabular}{lrrrrlrrrr}
\tophline
 & & \multicolumn{2}{c}{\textbf{Scenarios}} & & & \multicolumn{4}{c}{\textbf{Input group reported (\%)}} \\
\textbf{Family} & \textbf{Versions} & AR6 & retained & \textbf{Series} & \textbf{Base year} & Cap. & Eff. & Life. & O\&M \\
\middlehline
REMIND    &  9 / 12 &   323 &   248 &  6{,}127 & 2005              & 100 &  92 &  92 &  92 \\
MESSAGEix &  3 / 6  &   229 &   197 &  4{,}364 & 1990, 2010$^{*}$  & 100 &   0 &  97 &  97 \\
WITCH     &  1 / 5  &   176 &    86 &  2{,}288 & 2005              & 100 & 100 & 100 & 100 \\
IMAGE     &  3 / 4  &   156 &    78 &  2{,}148 & 1970              & 100 &  38 &   0 &  98 \\
IMACLIM   &  2 / 4  &    79 &    70 &  1{,}928 & 2001              & 100 &  99 &   0 & 100 \\
TIAM      &  2 / 9  &    96 &    52 &  1{,}919 & 2005$^{*}$        & 100 &   0 &   8 &   8 \\
POLES     &  2 / 6  &   139 &    81 &  1{,}781 & 2015$^{*}$        & 100 & 100 & 100 & 100 \\
GEM       &  1 / 2  &    52 &    41 &  1{,}361 & 2020$^{*}$        & 100 &   0 &   0 &   0 \\
AIM       & 11 / 18 &   468 &   213 &  1{,}021 & 2005, 2010$^{*}$  & 100 &  95 &   2 &  98 \\
GCAM      &  3 / 11 &   186 &    24 &     470 & 2010, 2015        & 100 &   0 &   0 &   0 \\
MARKAL    &  1 / 1  &    20 &    20 &      20 & 2001              & 100 & 100 & 100 & 100 \\
MERGE     &  1 / 1  &     3 &     3 &      18 & 2015              & 100 &   0 &   0 &   0 \\
C3IAM     &  1 / 2  &    14 &     3 &       9 & 2020$^{*}$        & 100 & 100 & 100 & 100 \\
LUT       &  2 / 3  &     4 &     3 &       7 & 2020$^{*}$        & 100 & 100 & 100 & 100 \\
DDPP      &  1 / 1  &     4 &     1 &       1 & 2020$^{*}$        & 100 & 100 & 100 & 100 \\
\middlehline
\textbf{All} & 43 / 85 & 1{,}949 & 1{,}120 & 23{,}462 & & & & & \\
\bottomhline
\end{tabular}

\vspace{2mm}
\noindent\footnotesize\emph{Retained versions:}
REMIND 1.7, REMIND 2.1, REMIND-Buildings 2.0, REMIND-H13 2.1, REMIND-MAgPIE 1.7-3.0, REMIND-MAgPIE 2.0-4.1, REMIND-MAgPIE 2.1-4.2, REMIND-MAgPIE 2.1-4.3, REMIND-Transport 2.1;
MESSAGEix-GLOBIOM\_1.0, MESSAGEix-GLOBIOM\_1.1, MESSAGEix-GLOBIOM\_GEI 1.0;
WITCH 5.0;
IMAGE 3.0, IMAGE 3.0.1, IMAGE 3.2;
IMACLIM 1.1, IMACLIM-NLU 1.0;
TIAM-ECN 1.1, TIAM-Grantham 3.2;
POLES ENGAGE, POLES GECO2019;
GEM-E3\_V2021;
AIM/CGE 2.1, AIM/CGE 2.2, AIM/CGE-Korea 2.1, AIM/Enduse-Japan 2.1, AIM/Hub-China 2.2, AIM/Hub-Global 2.0, AIM/Hub-India 2.2, AIM/Hub-Japan 2.1, AIM/Hub-Korea 2.0, AIM/Hub-Thailand 2.2, AIM/Hub-Vietnam 2.2;
GCAM 5.2, GCAM 5.3, GCAM-PR 5.3;
MARKAL-India 1.0;
MERGE-ETL 6.0;
C3IAM 1.0;
LUT-ESTM 1.0, LUT-ESTM 1.1;
DDPP Energy 1.0.
\end{table}

\begin{table}[h]
\footnotesize
\setlength{\tabcolsep}{3pt}
\renewcommand{\arraystretch}{1.1}
\caption{Variables of ML-IAM, their units, and coverage. Cov.\ is the share of the 23,462 retained series that report the variable in at least one year. Units are those of the AR6 database, with three exceptions: primary energy reported in EJ/yr is converted to PJ/yr, GDP|PPP reported in billion Int\$2010/yr is used as billion US\$2010/yr, and carbon prices whose unit does not name a currency are dropped.}
\label{tab:appendix_variables}
\begin{tabular}{r p{0.30\textwidth} l r @{\hspace{9pt}} r p{0.30\textwidth} l r}
\hline
 & Variable & Unit & Cov. & & Variable & Unit & Cov. \\
\hline
\multicolumn{8}{l}{\textbf{Inputs}} \\
\hline
 1  & Capital Cost|Electricity|Biomass|w/ CCS      & \$/kW    & 75\,\%  & 23 & Lifetime|Electricity|Geothermal           & years  & 50\,\% \\
 2  & Capital Cost|Electricity|Biomass|w/o CCS     & \$/kW    & 80\,\%  & 24 & Lifetime|Electricity|Hydro                & years  & 60\,\% \\
 3  & Capital Cost|Electricity|Coal|w/ CCS         & \$/kW    & 75\,\%  & 25 & Lifetime|Electricity|Nuclear              & years  & 60\,\% \\
 4  & Capital Cost|Electricity|Coal|w/o CCS        & \$/kW    & 80\,\%  & 26 & Lifetime|Electricity|Solar|PV             & years  & 60\,\% \\
 5  & Capital Cost|Electricity|Gas|w/ CCS          & \$/kW    & 77\,\%  & 27 & OM Cost|Fixed|Electricity|Biomass|w/ CCS  & \$/kW/yr & 56\,\% \\
 6  & Capital Cost|Electricity|Gas|w/o CCS         & \$/kW    & 75\,\%  & 28 & OM Cost|Fixed|Electricity|Biomass|w/o CCS & \$/kW/yr & 53\,\% \\
 7  & Capital Cost|Electricity|Geothermal          & \$/kW    & 63\,\%  & 29 & OM Cost|Fixed|Electricity|Coal|w/ CCS     & \$/kW/yr & 57\,\% \\
 8  & Capital Cost|Electricity|Hydro               & \$/kW    & 88\,\%  & 30 & OM Cost|Fixed|Electricity|Coal|w/o CCS    & \$/kW/yr & 63\,\% \\
 9  & Capital Cost|Electricity|Nuclear             & \$/kW    & 99\,\%  & 31 & OM Cost|Fixed|Electricity|Gas|w/ CCS      & \$/kW/yr & 57\,\% \\
10  & Capital Cost|Electricity|Solar|CSP           & \$/kW    & 69\,\%  & 32 & OM Cost|Fixed|Electricity|Gas|w/o CCS     & \$/kW/yr & 63\,\% \\
11  & Capital Cost|Electricity|Solar|PV            & \$/kW    & 100\,\% & 33 & OM Cost|Fixed|Electricity|Geothermal      & \$/kW/yr & 53\,\% \\
12  & Capital Cost|Electricity|Wind|Offshore       & \$/kW    & 69\,\%  & 34 & OM Cost|Fixed|Electricity|Hydro           & \$/kW/yr & 82\,\% \\
13  & Capital Cost|Electricity|Wind|Onshore        & \$/kW    & 69\,\%  & 35 & OM Cost|Fixed|Electricity|Nuclear         & \$/kW/yr & 81\,\% \\
14  & Capital Cost|Hydrogen|Electricity            & \$/kW    & 70\,\%  & 36 & OM Cost|Fixed|Electricity|Solar|CSP       & \$/kW/yr & 58\,\% \\
15  & Efficiency|Electricity|Biomass|w/ CCS        & \%       & 54\,\%  & 37 & OM Cost|Fixed|Electricity|Solar|PV        & \$/kW/yr & 80\,\% \\
16  & Efficiency|Electricity|Biomass|w/o CCS       & \%       & 53\,\%  & 38 & OM Cost|Fixed|Electricity|Wind|Offshore   & \$/kW/yr & 53\,\% \\
17  & Efficiency|Electricity|Coal|w/ CCS           & \%       & 56\,\%  & 39 & OM Cost|Fixed|Electricity|Wind|Onshore    & \$/kW/yr & 52\,\% \\
18  & Efficiency|Electricity|Coal|w/o CCS          & \%       & 54\,\%  & 40 & Population                                & million  & 92\,\% \\
19  & Efficiency|Electricity|Gas|w/ CCS            & \%       & 56\,\%  & 41 & Price|Carbon                              & \$/t CO$_2$ & 85\,\% \\
20  & Efficiency|Electricity|Gas|w/o CCS           & \%       & 54\,\%  & 42 & Yield|Cereal                              & t DM/ha/yr & 43\,\% \\
21  & GDP|MER                                      & bn \$/yr & 91\,\%  & 43 & Model Family                              & category & 100\,\% \\
22  & GDP|PPP                                      & bn \$/yr & 83\,\%  & 44 & Region                                    & category & 100\,\% \\
\hline
\multicolumn{8}{l}{\textbf{Outputs}} \\
\hline
 1  & Primary Energy|Coal      & PJ/yr & 100\,\% & 6 & Primary Energy|Nuclear & PJ/yr & 100\,\% \\
 2  & Primary Energy|Gas       & PJ/yr & 100\,\% & 7 & Emissions|CO2          & Mt CO$_2$/yr & 91\,\% \\
 3  & Primary Energy|Oil       & PJ/yr & 100\,\% & 8 & Emissions|CH4          & Mt CH$_4$/yr & 91\,\% \\
 4  & Primary Energy|Solar     & PJ/yr & 94\,\%  & 9 & Emissions|N2O          & kt N$_2$O/yr & 91\,\% \\
 5  & Primary Energy|Wind      & PJ/yr & 94\,\%  &   &                        &       &         \\
\hline
\end{tabular}

\vspace{2mm}
\noindent\footnotesize\emph{Units:} \$ = US\$2010; bn = billion; DM = dry matter. \emph{Abbreviations:} CCS = Carbon Capture and Storage; OM = Operations \& Maintenance; MER = Market Exchange Rate; PPP = Purchasing Power Parity.
\end{table}

\clearpage
\section{Architectures Not Considered}
\label{appendix:model_review}
The three models were chosen to span three formulations of the task under one tuning protocol, not to identify the best architecture within any formulation (Section~\ref{sec:method_selection}). This appendix explains why recent time-series and emulation architectures were not considered. Most are designed for regimes with characteristics opposite to those of Section~\ref{sec:method_task_char}, and were left out for that reason rather than on the basis of experiments.

\textbf{Long-horizon transformer models} (Autoformer \citep{wu2021autoformer}, FEDformer \citep{zhou2022fedformer}, Crossformer \citep{zhang2023crossformer}) are built to exploit seasonal patterns and long-range dependencies over context lengths of typically 96--720 steps. Neither exists in IAM series of at most 20 timesteps, which are driven by policy interventions rather than cyclical dynamics.

\textbf{Patch-based models} (PatchTST \citep{nie2022time}) decompose time series into local patches to capture recurring motifs, as in electricity demand or traffic flow. IAM trajectories instead exhibit structural shifts driven by discrete policy choices and technological transitions.

\textbf{Time-series foundation models} (TimesFM \citep{das2024decoder}, Chronos \citep{ansari2024chronos}, Moirai \citep{woo2024unified}, Sundial \citep{liu2025sundial}) offer zero-shot forecasting from a prior learned on large corpora of observed, largely periodic series, in which the past is informative about the future. IAM trajectories are instead counterfactual projections shaped by scenario assumptions, and their non-periodic, policy-driven shape (Section~\ref{sec:method_task_char}) is absent from such corpora. Using these models here would therefore mean fine-tuning them on this dataset, a study of transfer from a mismatched prior that lies outside the scope of a first feasibility demonstration.

\textbf{Gaussian-process emulators} are widely used in the climate and Earth-system emulation literature \citep{climatebench} when a simulator can be run only tens to hundreds of times, because they interpolate smoothly between few runs and return calibrated uncertainty. Their cost, however, scales cubically with the number of training points \citep{rasmussen2006gaussian}, and the present task has 186{,}000--282{,}000 training rows, depending on the model (Table~\ref{tab:split_breakdown}), well beyond the reach of exact inference. Sparse and variational approximations \citep{hensman2013gaussian} make training possible but forfeit much of what makes the method attractive in the small-ensemble regime.

\clearpage
\section{Training and Evaluation Details}
\label{appendix:training_details}
This appendix describes how each model produces a trajectory (Appendix~\ref{appendix:prediction_protocols}) and the settings of the final models (Appendix~\ref{appendix:model_settings}).
 
\subsection{Prediction Protocols}
\label{appendix:prediction_protocols}
 
All three emulators generate the outputs of a scenario from a window of that scenario's inputs. Let $\mathbf{u}_t$ denote the vector of scenario inputs at timestep $t$ (the 42 numerical variables of Table~\ref{tab:appendix_variables} after preprocessing, the model-family and region identifiers, and for the neural models the binary missingness indicators of Section~\ref{sec:method_model_setting}), and $\mathbf{y}_t$ the nine outputs. Timesteps are the reported timesteps of a series, five years apart for most series (Section~\ref{sec:method_model_setting}), and a trajectory is one (model, scenario, region) series. Each model was evaluated at two context lengths (2 and 3 timesteps) during hyperparameter search (Appendix~\ref{appendix:hpo}); the final models use the context lengths listed below.

\paragraph{XGBoost: recursive one-step prediction.}
XGBoost predicts $\mathbf{y}_t$ from the current inputs and the outputs of the two preceding timesteps,
\begin{equation}
    \hat{\mathbf{y}}_t = f_{\mathrm{XGB}}\!\left(\mathbf{u}_t, \mathbf{y}_{t-1}, \mathbf{y}_{t-2}\right),
\end{equation}
with one booster per output variable. Each booster's lag columns hold only its own output at the $L$ preceding timesteps, so the nine rollouts are independent of one another. XGBoost is trained teacher-forced, with the lagged outputs set to the reported IAM values, but deployed on its own predictions. The first predicted timestep is the fourth for either context length $L$, so that the two settings are compared on the same rows. Its lags are the reported values of the second and third timesteps; thereafter each lag is the model's own prediction once one exists for that timestep, so the reported third value is still used at the fourth and fifth timesteps. A gap in a reported output is filled for lag use with the last reported value before it. The first three timesteps of each series therefore receive no prediction. XGBoost is scored on the rollout rather than teacher-forced (one step ahead with reported lags), in every hyperparameter-search trial and for the final model (Appendix~\ref{appendix:hpo}).
 
\paragraph{LSTM: direct one-step regression from an input window.}
The LSTM maps the inputs of the three most recent timesteps onto the current outputs,
\begin{equation}
    \hat{\mathbf{y}}_t = f_{\mathrm{LSTM}}\!\left(\mathbf{u}_{t-2}, \mathbf{u}_{t-1}, \mathbf{u}_t\right),
\end{equation}
where the categorical identifiers in $\mathbf{u}$ enter through learned embeddings. No output values, reported or predicted, are given to the model at any point, so the LSTM is neither teacher-forced nor autoregressive. The hidden state is reset for each window. The first two timesteps of each trajectory receive no prediction. The sliding windows are aligned so that every context length predicts the same target rows (Appendix~\ref{appendix:hpo}).

\paragraph{TFT: direct multi-horizon forecast from an input window.}
The TFT forecasts 12 consecutive timesteps in a single pass from an encoder window of three preceding timesteps and a decoder window of the 12 timesteps to be predicted,
\begin{equation}
    \left(\hat{\mathbf{y}}_{t+1}, \ldots, \hat{\mathbf{y}}_{t+12}\right) = f_{\mathrm{TFT}}\!\left(\mathbf{u}_{t-2}, \ldots, \mathbf{u}_{t}; \mathbf{u}_{t+1}, \ldots, \mathbf{u}_{t+12}\right).
\end{equation}
All scenario inputs are treated as known-future covariates, since they are fixed by the scenario definition, and as for the LSTM no output values enter the model: the target series is deliberately excluded from the encoder inputs, and the per-target scale features that pytorch-forecasting attaches to each sample are constants under the global target normalizer used here (Appendix~\ref{appendix:model_settings}). The only other input is the relative time index that pytorch-forecasting adds to each window. At test time the decoder length is fixed at 12 timesteps and the encoder at three, so a trajectory of $T$ timesteps is covered by two windows: an \textit{early} window over its first 15 timesteps, predicting timesteps 4--15, and a \textit{late} window over its last 15 timesteps, predicting timesteps $T-11$ to $T$. Where the two windows overlap, their predictions are combined with a linear weight that fades from the early to the late window across the overlap. The first three timesteps of each trajectory receive no prediction, and trajectories shorter than 15 timesteps ($3+12$), which cannot fill one window, are excluded from the TFT evaluation. In training and validation the decoder length varies from 1 to 12 timesteps, so series with at least four timesteps are used.
 
\paragraph{Scored elements.}
All metrics are computed over target elements that were reported by the IAM (observed elements), on the timesteps each model predicts. The resulting counts differ by model for the reasons given above (Table~\ref{tab:split_breakdown}; see also Section~\ref{sec:results_eval}).
 
\subsection{Model Settings}
\label{appendix:model_settings}
 
The settings below are those of the final models; searched hyperparameters take the values selected by the procedure of Appendix~\ref{appendix:hpo} (Table~\ref{tab:hpo_search_space}), and all others are fixed defaults. Each model uses one configuration for all nine outputs. For LSTM and TFT this is inherent, since one network predicts all outputs from a shared representation. XGBoost fits one booster per output, so per-output configurations would be possible, but we keep a single one so that the search protocol is identical across models. All models were trained on NVIDIA RTX 2080~Ti GPUs; the LSTM and TFT final fits used eight GPUs with distributed data parallelism, and all inference ran on one GPU (Table~\ref{tab:phase_timings}).
 
\paragraph{Shared data handling.}
Series are split into training, validation, and test sets (80/10/10\,\%) with a fixed seed, so that all timesteps of a series fall in one split; the other regions of the same scenario are assigned independently (Section~\ref{sec:method_data_splitting}). The four smallest model families (C3IAM, DDPP, LUT, MERGE; 35 series in total) have no series in the validation set, and DDPP and MERGE none in the test set; among regions, R6ROWO (two series, both in the validation set) has none in the training or test set, and three countries (KAZ, PAK, TWN) none in the test set. Continuous inputs and targets are standardized with statistics fitted on the training split; for all three models this is one global scaler per variable (for the TFT, in place of pytorch-forecasting's per-window normalization, see below). The target statistics are computed from reported values for XGBoost and the LSTM. For the TFT they also include unreported elements, which are filled before the scaler is fitted; this affects only the network's internal scale, since its loss is in native units. Target elements that are not reported are excluded from the loss and from every metric through a per-element observation mask; rows in which no target is reported are dropped. Models are trained on the training split with early stopping on the validation split, and the checkpoint with the lowest validation loss is evaluated on the test split. The final model is not retrained on the combined training and validation data.
 
\paragraph{XGBoost.}
The XGBoost emulator consists of nine gradient-boosted tree ensembles (one per output variable, \texttt{xgboost} with the \texttt{hist} tree method on GPU), each trained on the rows in which its target is reported. Missing input values are handled by XGBoost's native sparsity-aware split finding, so no imputation or missingness indicators are used; the model-family and region identifiers enter as integer category codes. Each booster reads two lag features, its own output at the two preceding timesteps. Boosting uses a maximum of 1{,}500 rounds with early stopping on the validation RMSE, with a patience of 15 rounds during the search and 50 rounds for the final fit; the nine boosters of the final model stopped at 155--302 rounds. Searched hyperparameters are the tree depth, minimum child weight, $\gamma$, learning rate $\eta$, $L_1$ and $L_2$ regularization, row and column subsampling, and the context length (Table~\ref{tab:hpo_search_space}).

\paragraph{Linear baseline.}
The baseline is one ridge regression per output (scikit-learn, default penalty $\alpha = 1$, not tuned) on the XGBoost feature set, including the lag features, with the same splits and the same autoregressive rollout. Because a linear fit cannot use missing values or integer category codes, missing inputs are filled with training-set medians and the model-family and region identifiers are one-hot encoded (154 columns in total, of which each output's regression uses 138, since it reads only its own two lag columns). A ridge penalty is needed because unpenalized least squares diverged in the rollout; validation rollout $R^2$ is unchanged to three decimals for $\alpha$ between $10^{-6}$ and $10^{3}$. Training takes 17\,s on CPU. Its test-set $R^2$ of 0.710 flatters a linear emulator (Figs.~S4 and S8): the predictions saturate at high values (coal, nuclear), turn negative for small series (solar, wind, CO$_2$), and drift away from the IAM paths as the rollout feeds its own errors back through the lag features. On the C3 World trajectories of Fig.~S5 the per-target average $R^2$ is 0.31 (coal and gas below zero) against 0.93 for XGBoost on the same rows; the pooled $R^2$ on those rows, 0.67, is dominated by the large-magnitude series and hides this.
 
\paragraph{LSTM.}
The LSTM (PyTorch Lightning) consists of an embedding layer for each categorical identifier (16-dimensional), a unidirectional three-layer LSTM with 328 hidden units, and a two-layer dense head (48 units, ReLU) that maps the final hidden state to the nine outputs. Dropout of 0.282 is applied between LSTM layers and 0.083 in the dense head. The loss is the mean squared error over observed target elements in standardized units. Training uses Adam with a learning rate of $3.6\times10^{-4}$, weight decay of $4.2\times10^{-8}$ (effectively none), batch size 128, gradient-norm clipping at 1.0, and a maximum of 100 epochs, with early stopping on the validation loss (patience 3 epochs during stage-1 search, 20 epochs for stage-2 refits and the final fit); the final model reached the 100-epoch budget with its validation loss still decreasing (lowest at the last epoch). Missing continuous inputs are imputed with the training-split median of the same region scale, and a binary indicator per input marks the imputed elements. Windows are built from consecutive reported timesteps of a series, whatever their spacing (Section~\ref{sec:method_model_setting}).
 
\paragraph{TFT.}
The TFT (pytorch-forecasting) uses a hidden size of 496, one LSTM layer in the encoder and decoder, four attention heads, a hidden continuous size of 8, and dropout of 0.019. The encoder length is three timesteps and the prediction length 12. All inputs are declared as time-varying known covariates and the model-family and region identifiers as static categoricals; the targets are not included among the encoder inputs. Targets are normalized with a single global standard scaler per variable fitted on the training split rather than with pytorch-forecasting's default per-window (encoder) normalization, which in our preliminary runs produced unstable scales for series near zero. The loss is a masked root mean squared error summed over the nine targets, which ignores unobserved elements. Following pytorch-forecasting's convention, it is evaluated after the network output is mapped back through the target scaler, so the loss, the search scores of Table~\ref{tab:best_params} (computed without the mask, Appendix~\ref{appendix:hpo}) and the learning curves of Fig.~\ref{fig:learning_curves} are all in native units and are dominated by the large-magnitude energy and CO$_2$ series. Training uses Adam with a learning rate of $1.79\times10^{-4}$, batch size 64, gradient-norm clipping at 0.1, and a maximum of 100 epochs, with early stopping on the validation loss (patience 8 epochs during search, 20 for the final fit); the final model reached the 100-epoch budget without triggering early stopping, the lowest validation loss occurring at epoch 95 (epochs numbered from 1, as in Fig.~\ref{fig:learning_curves}). Missing inputs are imputed and flagged as for the LSTM.
 
\paragraph{Learned embeddings.}
A PCA projection of the trained embeddings (Fig.~S12) recovers some known structure: in both models most of the large energy-system partial-equilibrium families are each other's nearest neighbors, and the EU sits next to its R10 macro-region, but the CGE and intertemporal-optimization families do not group and most other large economies are not placed near their own region.
 
\clearpage
\section{Hyperparameter Optimization Details}
\label{appendix:hpo}
 
The three architectures have different hyperparameters, so their search spaces differ, but all three run the same two-stage random-search protocol and differ only in the per-trial training budget.
 
\subsection{Protocol.}
\begin{enumerate}
    \item \textbf{Stage 1 -- screening.} 50 configurations are drawn from the model's search space (Table~\ref{tab:hpo_search_space}) with a fixed random seed. Each configuration is trained under a reduced budget (XGBoost: 300 boosting rounds; LSTM: 20 epochs; TFT: 25 epochs) and scored on the validation split.
    \item \textbf{Stage 2 -- refit.} The eight best stage-1 configurations are retrained under the full training budget (1{,}500 rounds, 100 epochs, and 60 epochs respectively, each with early stopping) and re-scored, so that a configuration is judged on where it converges rather than on how fast it starts. The best stage-2 configuration is selected.
    \item \textbf{Final fit.} The selected configuration is trained once more on the training split with early stopping on the validation split, and this model is evaluated on the test split (Appendix~\ref{appendix:model_settings}). Table~\ref{tab:best_params} gives its validation score and the epoch or boosting round at which early stopping selected it.
\end{enumerate}
Two details concern how trials are scored. First, XGBoost trials are scored on the autoregressive rollout of Appendix~\ref{appendix:prediction_protocols}, not on one-step predictions from reported lags: a model can track the truth one step ahead and still compound its own errors (the linear baseline scores $R^2 = 0.988$ one step ahead from reported lags and 0.710 on the rollout), and XGBoost is the only model for which the two differ. Second, the context length (2 or 3 timesteps) is a searched hyperparameter for every model, and both settings are scored on the same target rows (Appendix~\ref{appendix:prediction_protocols}). For the TFT, $L=2$ is additionally scored on each series' third timestep; its best configurations at $L=3$ and $L=2$ differ by 0.5\% in validation loss, so the comparison does not strongly favor either. The TFT search was run before the observation mask was applied in the loss and with the observed-value flags among the inputs; the final model is the selected configuration refitted with the masked loss and without those inputs, and the search was not repeated, so its validation score in Table~\ref{tab:best_params} is not comparable with the final-fit loss in Fig.~\ref{fig:learning_curves}.

\begin{table}[h]
\small
\centering
\caption{Search spaces of the stage-1 random search (50 trials per model). $\mathcal{U}$: uniform; $\log\mathcal{U}$: log-uniform; integer ranges are sampled log-uniformly and rounded to the stated multiple. The context length is searched over the same two values for all models.}
\label{tab:hpo_search_space}
\begin{tabular}{llll}
\tophline
\textbf{Model} & \textbf{Hyperparameter} & \textbf{Distribution} & \textbf{Selected} \\
\middlehline
\multirow{9}{*}{XGBoost}
 & Max depth & integer $\mathcal{U}[4, 18]$ & 17 \\
 & Min child weight & integer $\log\mathcal{U}[1, 40]$ & 3 \\
 & $\gamma$ & $\mathcal{U}[0, 0.5]$ & 0.060 \\
 & Learning rate $\eta$ & $\log\mathcal{U}[0.01, 0.5]$ & 0.033 \\
 & $L_1$ regularization $\alpha$ & $\log\mathcal{U}[10^{-8}, 10]$ & $6.1\times10^{-4}$ \\
 & $L_2$ regularization $\lambda$ & $\log\mathcal{U}[10^{-2}, 100]$ & 24.1 \\
 & Row subsample & $\mathcal{U}[0.6, 1.0]$ & 0.81 \\
 & Column subsample & $\mathcal{U}[0.6, 1.0]$ & 0.90 \\
 & Context length (lags) & $\{2, 3\}$ & 2 \\
\middlehline
\multirow{10}{*}{LSTM}
 & Hidden size & integer $\log\mathcal{U}[32, 512]$, multiple of 8 & 328 \\
 & Layers & $\{1, 2, 3\}$ & 3 \\
 & LSTM dropout & $\mathcal{U}[0, 0.4]$ & 0.282 \\
 & Dense hidden size & integer $\log\mathcal{U}[32, 256]$, multiple of 8 & 48 \\
 & Dense dropout & $\mathcal{U}[0, 0.3]$ & 0.083 \\
 & Learning rate & $\log\mathcal{U}[10^{-4}, 5\times10^{-2}]$ & $3.6\times10^{-4}$ \\
 & Batch size & $\{32, 64, 128, 256\}$ & 128 \\
 & Weight decay & $\log\mathcal{U}[10^{-8}, 10^{-3}]$ & $4.2\times10^{-8}$ \\
 & Embedding dimension & integer $\log\mathcal{U}[4, 64]$, multiple of 4 & 16 \\
 & Context length (window) & $\{2, 3\}$ & 3 \\
\middlehline
\multirow{5}{*}{TFT}
 & Hidden size & integer $\log\mathcal{U}[64, 512]$, multiple of 8 & 496 \\
 & LSTM layers & $\{1, 2, 3\}$ & 1 \\
 & Dropout & $\mathcal{U}[0, 0.5]$ & 0.019 \\
 & Learning rate & $\log\mathcal{U}[10^{-4}, 3\times10^{-2}]$ & $1.79\times10^{-4}$ \\
 & Context length (encoder) & $\{2, 3\}$ & 3 \\
\bottomhline
\end{tabular}
\end{table}

\begin{table}[h]
\small
\centering
\caption{Outcome of the two-stage search: validation score of the selected configuration after the stage-2 refit, and the epoch or boosting round at which early stopping selected it.}
\label{tab:best_params}
\begin{tabular}{lll}
\tophline
\textbf{Model} & \textbf{Stage-2 validation score} & \textbf{Stage-2 budget} \\
\middlehline
XGBoost & RMSE $= 0.168$ (rollout, standardized) & 226 rounds \\
LSTM    & MSE $= 0.0228$ (standardized) & epoch 87 \\
TFT     & RMSE $= 8677$ (native units, summed) & epoch 59 \\
\bottomhline
\end{tabular}
\end{table}

\subsection{Search-budget sensitivity.}
\begin{figure}[htbp]
    \centering
    \includegraphics[width=\linewidth]{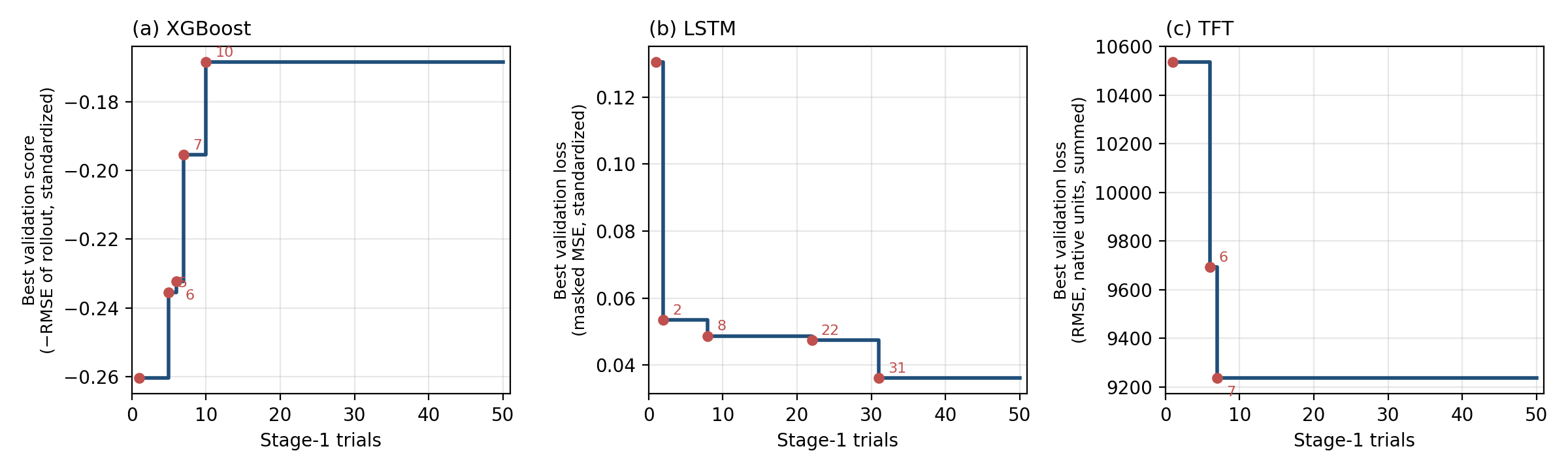}
    \caption{Search-budget sensitivity of the stage-1 random search: best validation score found as a function of the number of trials, for (a) XGBoost (negative validation RMSE of the autoregressive rollout, standardized units), (b) LSTM (masked validation MSE, standardized units), and (c) TFT (validation RMSE summed over targets, native units). Markers indicate the trials at which the best score improved.}
    \label{fig:hpo_budget}
\end{figure}

Figure~\ref{fig:hpo_budget} shows, for each model, the best validation score found after each stage-1 trial. The XGBoost curve last improves at trial 10, the LSTM curve at trial 31, and the TFT curve at trial 7, so the remaining trials did not change the selected configuration and a larger budget would be unlikely to. For XGBoost the trial-10 configuration also won stage 2.

\subsection{Computational cost.}
Table~\ref{tab:phase_timings} lists the wall-clock time of every pipeline phase for each model, as plotted in Fig.~\ref{fig:computational_cost}. The TFT stage-1 search ran as three shards on three machines; the table caption explains how their times are reported.

\begin{table}[htbp]
\small
\centering
\caption{Wall-clock time per pipeline phase on NVIDIA RTX 2080~Ti GPUs. Preprocessing runs on CPU and is shared by all models. Search times cover stage~1 (50 trials) and stage~2 (8 refits). For TFT, stage~1 ran as three shards on three machines with eight, seven, and eight GPUs; ``active'' is each shard's running time excluding the idle gap after a failed worker, ``single-machine equivalent'' is the sum of the three, and ``wall clock'' is first shard start to last shard end. Inference is end-to-end (data preparation, model loading, prediction, metrics) over the test trajectories each model predicts. Search totals quoted in the text are computed from unrounded stage times.}
\label{tab:phase_timings}
\begin{tabular}{llrrl}
\tophline
\textbf{Model} & \textbf{Phase} & \textbf{Time} & \textbf{GPUs} & \textbf{Notes} \\
\middlehline
All & Preprocess & 40\,s & CPU & 352{,}898 rows \\
\middlehline
\multirow{4}{*}{XGBoost}
 & Search, stage 1 & 4.8\,min & 8 & 50 trials $\times$ 300 rounds \\
 & Search, stage 2 & 1.3\,min & 8 & 8 refits, $\leq$1{,}500 rounds \\
 & Final fit & 51\,s & 1 & 155--302 rounds per output \\
 & Inference & 14\,s & 1 & 2{,}346 trajectories; rollout alone 0.6\,s \\
\middlehline
\multirow{4}{*}{LSTM}
 & Search, stage 1 & 2.7\,h & 8 & 50 trials $\times$ 20 epochs \\
 & Search, stage 2 & 4.9\,h & 8 & 8 refits, $\leq$100 epochs \\
 & Final fit & 17.5\,min & 8 & 100 epochs \\
 & Inference & 21\,s & 1 & 2{,}346 trajectories \\
\middlehline
\multirow{7}{*}{TFT}
 & Search, stage 1, shard 0 (active) & 64.2\,h & 8 & 17 trials $\times$ 25 epochs \\
 & Search, stage 1, shard 1 (active) & 98.9\,h & 7 & 17 trials; resumed once \\
 & Search, stage 1, shard 2 (active) & 89.9\,h & 8 & 16 trials; resumed once \\
 & Search, stage 1, single-machine equivalent & 253.0\,h & 8 & sum of shards (wall clock 118.1\,h on 23 GPUs) \\
 & Search, stage 2 & 67.0\,h & 8 & 8 refits, $\leq$60 epochs \\
 & Final fit & 17.6\,h & 8 & 100 epochs, $\approx$10.5\,min per epoch \\
 & Inference & 85\,s & 1 & 1{,}738 trajectories; model $\approx$16\,s, window blending $\approx$44\,s \\
\bottomhline
\end{tabular}
\end{table}

\clearpage
\section{SHAP Analysis Details}
\label{appendix:shap_details}
 
For XGBoost, we use TreeExplainer~\citep{lundberg2020local}, which computes exact SHAP values for tree ensembles. For LSTM and TFT, attributions are expected gradients~\citep{erion2021improving}, computed with a batched implementation of the estimator for a seeded random sample of 500 input windows against a background of 200 further windows drawn from the same subset, with 400 samples per window; only continuous inputs are attributed, and each window keeps its own categorical inputs fixed.

All three analyses use the same test subset: the R10 regions, from which 300 of the 513 (Model, Scenario) groups are drawn at random with a fixed seed, keeping all timesteps of their series. The unit explained is one (series, timestep) row for XGBoost, one three-step window for the LSTM, and one 15-step window (three encoder and 12 decoder steps) for the TFT; for XGBoost every row of the subset is explained except those with an unreported lag. Attributions are computed on the test inputs as reported by the IAMs; for XGBoost this means the lag features hold reported rather than predicted values. For every model, each output's SHAP values are divided by the sum over features of that output's mean absolute attribution, so that all panels of Fig.~\ref{fig:shap_CO2_all} and Figs.~S9--S11 share one scale; rankings within a panel do not depend on this scaling. The beeswarm plots show the eight features with the largest mean absolute attribution per output.
 
For XGBoost, each timestep is predicted independently, so an attribution is the contribution of a feature, including a lag feature, to the prediction at that timestep. For the LSTM, attributions cover the three input timesteps of each window and are summed over them. For the TFT, the explained quantity is the forecast over the decoder horizon, computed separately for each output; attributions cover the continuous inputs of all 15 timesteps and are summed over time, and each window keeps its own decoder length and target scale fixed.

\clearpage
\section{Extended Results}
\label{appendix:extended_results}

\subsection{Accuracy by regional scale}
\label{appendix:by_scale}
 
Table~\ref{tab:performance_summary} reports the test-set $R^2$ of each model, and of the linear baseline, within each regional scale. The baseline falls from 0.710 overall to 0.38 at the World level and 0.43--0.51 at the R10 and country scales. TFT retains $R^2 \geq 0.95$ at every scale, and XGBoost and LSTM at every scale but World (0.938 and 0.919).
 
\begin{table}[h]
\small
\centering
\caption{Test-set $R^2$ by model and regional scale, including the linear baseline (ridge regression with XGBoost's features and rollout). ``All'' is the unweighted mean of the nine per-target $R^2$ values; scale columns give the same mean restricted to that scale. $N$ is the number of scored target elements (Section~\ref{sec:method_protocol}).}
\label{tab:performance_summary}
\begin{tabular}{lccccccc}
\tophline
\textbf{Model} & \textbf{All} & World & R5 & R6 & R10 & ISO3 & $N$ \\
\middlehline
Linear baseline & 0.710 & 0.375 & 0.588 & 0.621 & 0.426 & 0.506 & 246{,}125 \\
XGBoost & 0.975 & 0.938 & \textbf{0.966} & 0.973 & 0.960 & 0.966 & 246{,}125 \\
LSTM    & 0.971 & 0.919 & 0.956 & 0.971 & 0.954 & 0.959 & 266{,}356 \\
TFT     & \textbf{0.985} & \textbf{0.961} & 0.965 & \textbf{0.990} & \textbf{0.982} & \textbf{0.982} & 207{,}793 \\
\bottomhline
\end{tabular}
\end{table}

\subsection{Per-target accuracy}
 
Table~\ref{tab:r2_results} reports the test-set $R^2$ of each model for every output variable. Extended visualizations, including scatter plots (Figures S1--S3), temporal trajectories (Figures S5--S7), and SHAP analysis (Figures S9--S11), are provided in the Supplementary Information.
 
\begin{table}[h]
\centering
\caption{Test-set $R^2$ of ML model predictions versus original IAM outputs for each target variable, computed over observed elements only. ``Per-target average'' is the unweighted mean of the nine values above it; ``Pooled'' is the $R^2$ over all observed test elements of all targets. The range row is the difference between the best and worst target of each model.}
\label{tab:r2_results}
\begin{tabular}{lccc}
\tophline
\textbf{Output Variable} & \textbf{XGBoost} & \textbf{LSTM} & \textbf{TFT} \\
\middlehline
Primary Energy|Coal     & 0.965 & 0.941 & 0.976 \\
Primary Energy|Gas      & 0.977 & 0.977 & 0.989 \\
Primary Energy|Oil      & 0.982 & 0.981 & 0.988 \\
Primary Energy|Solar    & 0.986 & 0.977 & 0.988 \\
Primary Energy|Wind     & 0.972 & 0.970 & 0.979 \\
Primary Energy|Nuclear  & 0.951 & 0.953 & 0.980 \\
Emissions|CO2           & 0.967 & 0.973 & 0.984 \\
Emissions|CH4           & 0.986 & 0.983 & 0.986 \\
Emissions|N2O           & 0.992 & 0.979 & 0.992 \\
\middlehline
\textbf{Per-target average} & \textbf{0.975} & \textbf{0.971} & \textbf{0.985} \\
\textbf{Pooled}             & \textbf{0.979} & \textbf{0.973} & \textbf{0.987} \\
Range (best $-$ worst)      & 0.042 & 0.042 & 0.016 \\
\bottomhline
\end{tabular}
\end{table}

\subsection{Generalization across splits}
\label{appendix:split_breakdown}
 
Table~\ref{tab:split_breakdown} reports the same metrics on the training, validation, and test splits, each evaluated with the model's deployment protocol (Appendix~\ref{appendix:prediction_protocols}) rather than the teacher-forced loss monitored during training. Train--test gaps in $R^2$ are 0.015 (XGBoost), 0.014 (LSTM), and 0.012 (TFT), and validation and test agree to within 0.006 for every model, so the validation-based selection of hyperparameters and checkpoints transfers to held-out series. RMSE and MAE are pooled over all targets in native units, so they are dominated by the large-magnitude energy and CO$_2$ series and show consistency across splits, not absolute accuracy.

\begin{table}[h]
\small
\centering
\caption{Performance of each model on the training, validation, and test splits under its deployment-time evaluation protocol. Rows are scenario--timestep rows with at least one prediction; $N$ is the number of observed target elements scored. RMSE and MAE are pooled over all targets in native units (see text).}
\label{tab:split_breakdown}
\begin{tabular}{llrrccc}
\tophline
\textbf{Model} & \textbf{Split} & \textbf{Rows} & $N$ & \textbf{$R^2$ (per-target avg)} & \textbf{RMSE} & \textbf{MAE} \\
\middlehline
\multirow{3}{*}{XGBoost}
 & Train      & 225{,}436 & 1{,}957{,}967 & 0.990 & 2109 &  700 \\
 & Validation &  27{,}915 &   242{,}421 & 0.970 & 3509 &  920 \\
 & Test       &  28{,}360 &   246{,}125 & 0.975 & 3402 &  900 \\
\middlehline
\multirow{3}{*}{LSTM}
 & Train      & 281{,}740 & 2{,}119{,}678 & 0.984 & 2665 & 1036 \\
 & Validation &  34{,}947 &   262{,}581 & 0.972 & 3226 & 1135 \\
 & Test       &  35{,}398 &   266{,}356 & 0.971 & 3842 & 1157 \\
\middlehline
\multirow{3}{*}{TFT}
 & Train      & 185{,}985 & 1{,}628{,}947 & 0.996 &  921 &  329 \\
 & Validation &  22{,}694 &   199{,}162 & 0.981 & 2463 &  567 \\
 & Test       &  23{,}801 &   207{,}793 & 0.985 & 2764 &  566 \\
\bottomhline
\end{tabular}
\end{table}

\subsection{Counterparts of test series in the training set}
\label{appendix:counterparts}

Because the split is by series (Section~\ref{sec:method_data_splitting}), the training set holds other regions of most test scenarios and, for scenarios from multi-model protocols, the same scenario name run by other IAMs. Table~\ref{tab:counterparts} counts these counterparts and compares the test $R^2$ of the series that have one with that of the series that do not. At the series level, a single one of the 2,346 test series has none of the three. The World aggregate makes no difference to any model, so the summation constraint between a region and its aggregate is not exploited. A same-name counterpart matters little for any model: the gaps are 0.011 (XGBoost), 0.015 (LSTM), and 0.004 (TFT). The two subsets also differ in composition, since protocol scenarios are reported mainly at aggregate scales and by the large families, so these gaps are upper bounds on what a counterpart contributes.

\begin{table}[h]
\small
\centering
\caption{Counterparts of the test scenarios in the training or validation split, and test-set $R^2$ (per-target average) on the test series with and without such a counterpart. The World row counts the 854 test scenarios with a non-World test series and scores non-World series only. Too few test scenarios lack another of their regions for a comparison.}
\label{tab:counterparts}
\begin{tabular}{lrccc}
\tophline
 & \textbf{Test scenarios} & \multicolumn{3}{c}{$R^2$ with / without counterpart} \\
\textbf{Counterpart in training or validation} & \textbf{with one} & \textbf{XGBoost} & \textbf{LSTM} & \textbf{TFT} \\
\middlehline
Another region of the same scenario         & 845 of 866 (98\,\%) & -- & -- & -- \\
The World aggregate of the same scenario    & 687 of 854 (80\,\%) & 0.970 / 0.966 & 0.964 / 0.962 & 0.978 / 0.982 \\
The same scenario name run by another IAM   & 495 of 866 (57\,\%) & 0.981 / 0.970 & 0.978 / 0.963 & 0.987 / 0.983 \\
\bottomhline
\end{tabular}
\end{table}

\subsection{Learning curves}
\label{appendix:learning_curves}
 
\begin{figure}[h]
    \centering
    \includegraphics[width=\linewidth]{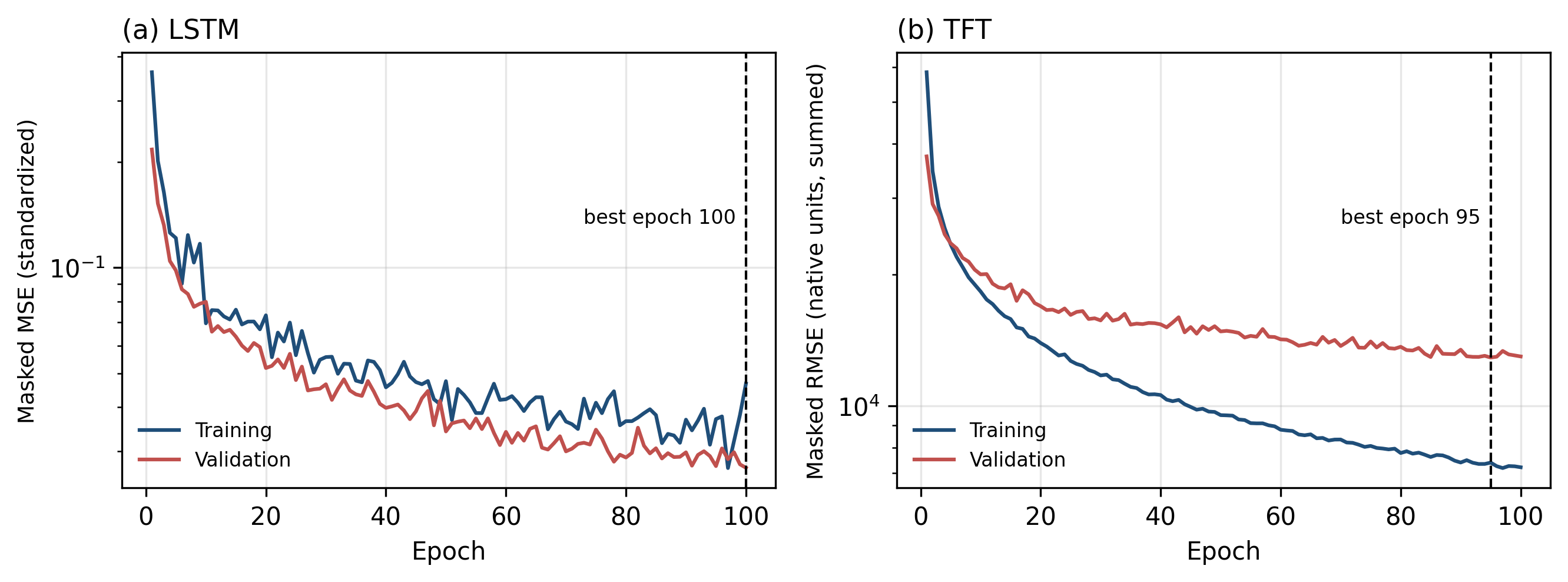}
    \caption{Learning curves of the final LSTM and TFT models: training and validation loss per epoch for (a) LSTM (masked MSE on standardized targets) and (b) TFT (masked RMSE, summed over targets in native units), on a log scale. Dashed lines mark the epoch with the lowest validation loss, whose checkpoint is the final model.}
    \label{fig:learning_curves}
\end{figure}

Figure~\ref{fig:learning_curves} shows the training and validation loss of the final LSTM and TFT models by epoch. Neither model triggered early stopping within the 100-epoch budget. The LSTM validation loss falls from 0.217 to 0.027 and tracks the training loss throughout, reaching its minimum at the last epoch. The TFT validation loss falls from 37{,}435 to 12{,}903 at epoch 95 and is flat thereafter; its training loss falls faster, to 7{,}217, so a gap opens within the first 10 epochs and widens throughout training, but the validation loss does not rise, consistent with the small train--test gap in Table~\ref{tab:split_breakdown}.
 
\clearpage

\clearpage
\section{ML-IAM Emulation Viewer}
\label{appendix:viewer_section}

\begin{figure}[p]
\centering
\includegraphics[width=\linewidth,height=0.85\textheight,keepaspectratio]{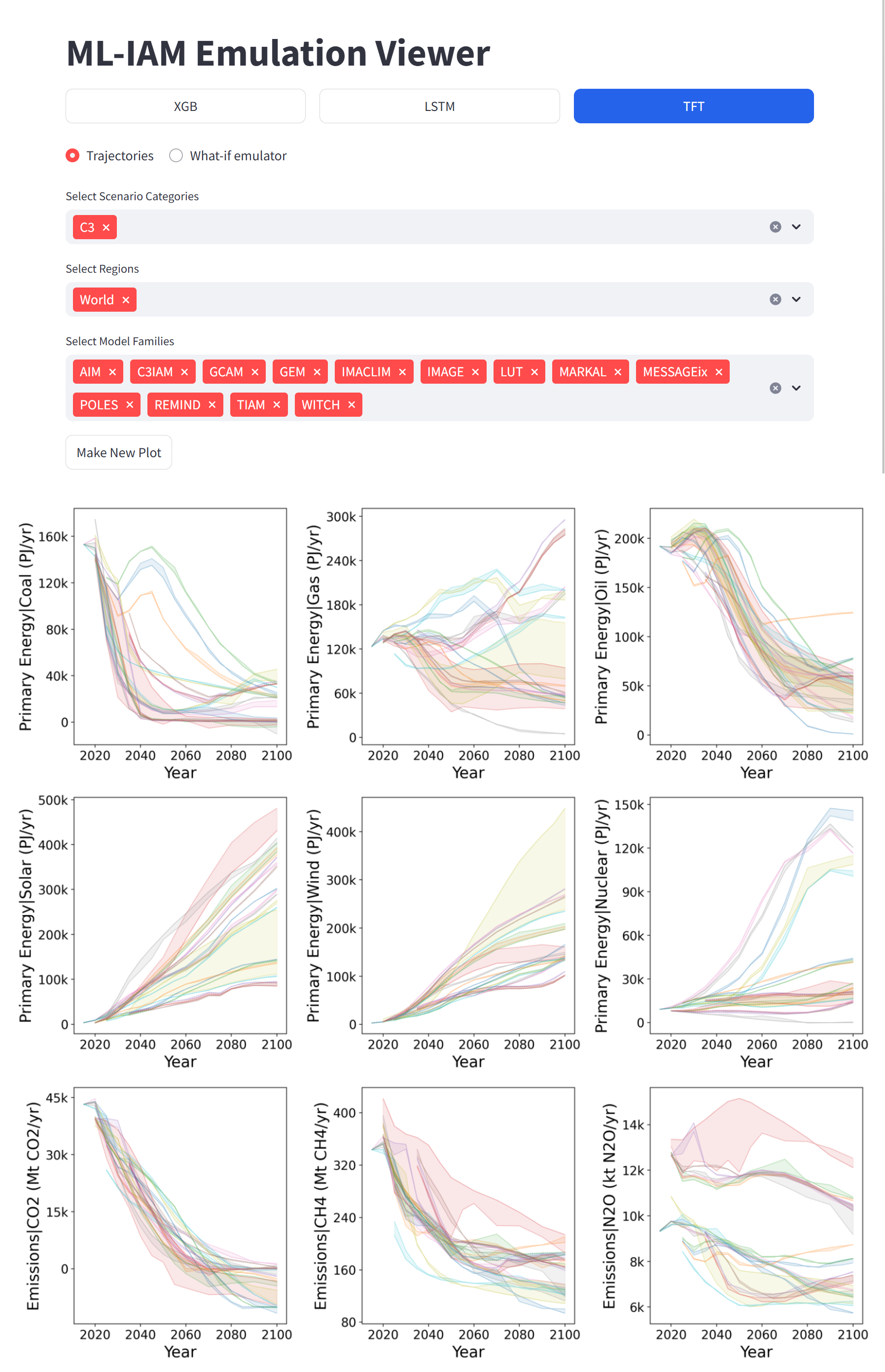}
\caption{Trajectories view of the Emulation Viewer. Top: model, view, and filter selection. Bottom: emulator predictions against the original IAM trajectories for the selected C3 scenarios in the World region.}
\label{appendix:viewer}
\end{figure}

\begin{figure}[p]
\centering
\includegraphics[width=\linewidth,height=0.85\textheight,keepaspectratio]{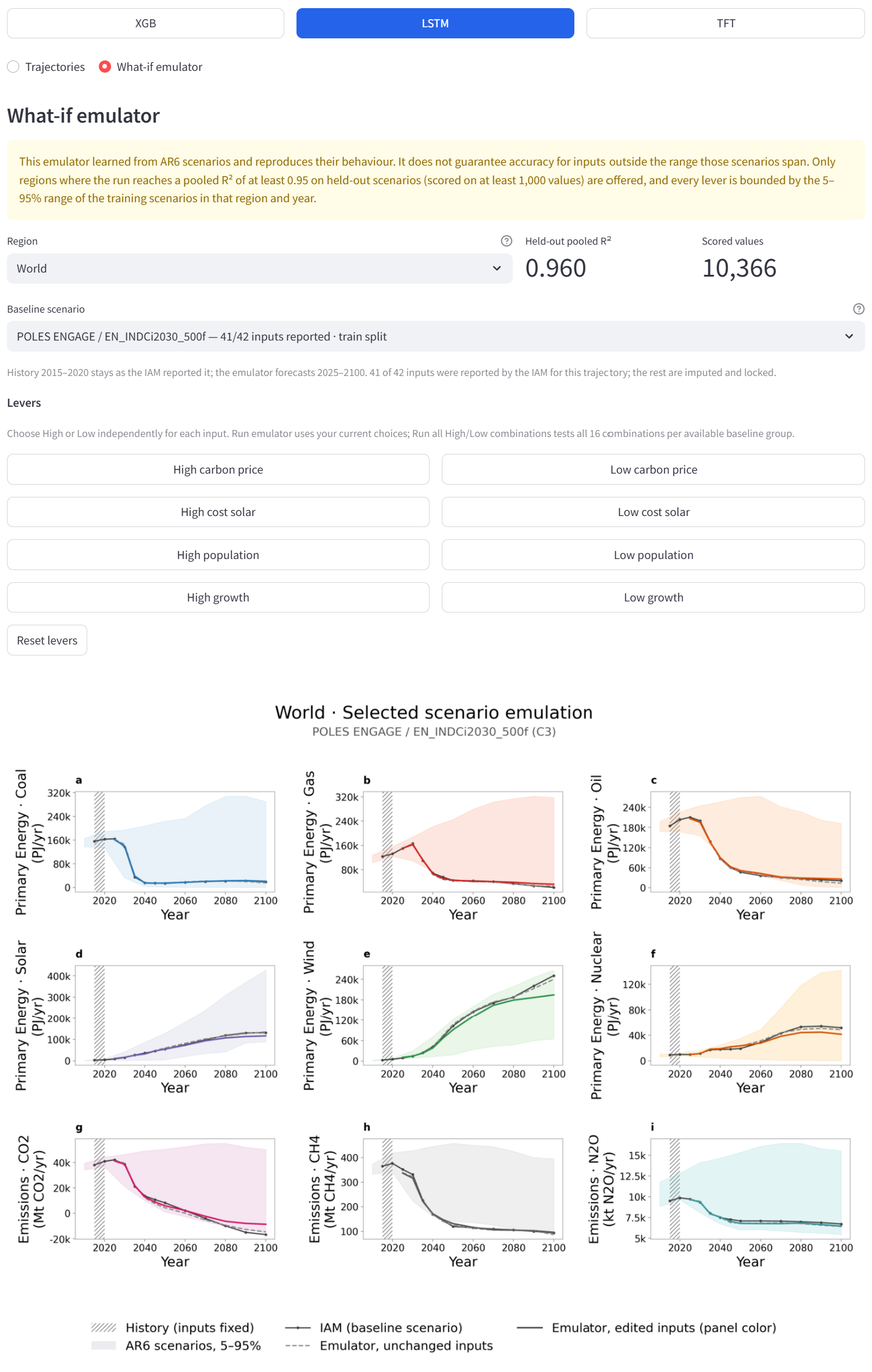}
\caption{What-if view of the Emulation Viewer. Top: region and baseline scenario selection with the High/Low presets. Bottom: emulated trajectories for the edited inputs next to the baseline IAM trajectory and the emulator's forecast of the unchanged inputs, with the 5th--95th percentile range of AR6 scenarios in the region shaded.}
\label{appendix:viewer_whatif}
\end{figure}

The trained emulators are available as a web application at \url{https://mliam.dev/}. The Trajectories view (Fig.~\ref{appendix:viewer}) plots the test-set predictions against the original IAM trajectories for user-selected scenario categories, regions, and IAM families. The What-if view (Fig.~\ref{appendix:viewer_whatif}) runs a trained emulator live: starting from the inputs of an AR6 scenario in a chosen region, the user moves inputs such as carbon price, technology costs, population, and GDP within the range of the training scenarios and obtains the emulated trajectories in seconds; results can be downloaded as CSV.

\clearpage 

\noappendix       







\authorcontribution{
YS, CL, JW, AO, and HM designed the study. YS, JH, and KP performed the literature review. JH, JW, MC, BK, and HM contributed to the data curation and processing. CL, EK, and JM contribute to the development of the ML models. JH, EK, and HM contributed to the visualizations. YS developed the model code, conducted the analysis, and wrote the original manuscript draft with contributions from all co-authors.
} 

\competinginterests{No competing interests.} 


\begin{acknowledgements}
This research was supported by the Korea Environmental Industry \& Technology Institute (KEITI) through Project for developing an observation-based GHG emissions geospatial information map, funded by Korea Ministry of Environment (MOE) (RS-2023-00232066).

The authors used generative AI tools to assist with code development and to improve the readability and clarity of the manuscript. All AI-generated content was reviewed and edited by the authors, who take full responsibility for the final publication.
\end{acknowledgements}







\bibliographystyle{copernicus}
\bibliography{references.bib}

\end{document}